\documentclass[reprint,aps,prc,groupedaddress]{revtex4-1}

\usepackage{graphicx}
\usepackage{dcolumn}
\usepackage{bm}
\usepackage{amsmath, amssymb}
\usepackage{threeparttable, multirow, booktabs}
\usepackage{longtable}
\usepackage{color}
\usepackage{mathrsfs}

\begin{document}


\title{$\alpha$ Decay Half-life Estimation and Uncertainty Analysis}


\author{Boshuai Cai}
\thanks{these authors contribute equally to this work}
\affiliation{Sino-French Institute of Nuclear Engineering and Technology, Sun Yat-Sen University, Zhuhai, 519082, Guangdong, China}
\author{Guangshang Chen}
\thanks{these authors contribute equally to this work}
\affiliation{Sino-French Institute of Nuclear Engineering and Technology, Sun Yat-Sen University, Zhuhai, 519082, Guangdong, China}
\author{Jiongyu Xu}
\affiliation{Sino-French Institute of Nuclear Engineering and Technology, Sun Yat-Sen University, Zhuhai, 519082, Guangdong, China}
\author{Cenxi Yuan}
\email{yuancx@mail.sysu.edu.cn}
\affiliation{Sino-French Institute of Nuclear Engineering and Technology, Sun Yat-Sen University, Zhuhai, 519082, Guangdong, China}
\author{Chong Qi}
\affiliation{KTH Royal Institute of Technology, Albanova University Center, SE-10691, Stockholm, Sweden}
\author{Yuan Yao}
\affiliation{College of Nuclear Science and Technology, Beijing Normal University, Beijing 100875, China}


\date{\today}

\begin{abstract}
\begin{description}
\item[Background]  $\alpha$ decay is one of the most important decay modes of atomic nuclei. The half-life of $\alpha$ decay provides valuable information for nuclear structure study. Many theoretical models and empirical formulas have been suggested to describe the half-life of $\alpha$ decay as a function of decay energy ($Q_{\alpha}$), atomic number (Z), nucleon number (A), and other related. However, the analysis of theoretical uncertainty is rarely done for those $\alpha$ decay models.
\item[Purpose]     We aim to perform a systematic and detailed study on the theoretical uncertainty of existing $\alpha$ decay formulas based on statistical methods.
\item[Methods]     The non-parametric bootstrap method is used to evaluate the uncertainties of two $\alpha$ decay formulas, the universal decay law (UDL) and the new Geiger-Nuttall law (NGNL). Such a method can simultaneously obtain the uncertainty of each parameter, the correlation between each pair of parameters, and the total, statistical, and systematic uncertainties of each formula. Both even-even (ee) nuclei and odd-A (oA) nuclei are used in the analysis. The collected data are separated into three parts: ee nuclei, oA nuclei without spin or parity change (oA\_nc), and oA nuclei with spin and/or parity change (oA\_c). Based on the residues between observed data and corresponding calculations, the statistical and systematic uncertainties are decomposed from the total uncertainty, from which one can clarify the effects from the shell structure, pairing, and angular momentum change on describing $\alpha$ decay half-life.
\item[Results]     If $N > 126$ and $N \leqslant 126$ nuclei are considered together, the systematic uncertainty of residues between observed and predicted half-lives are larger than if those groups are considered separately. Without shell correction term, a much larger systematic uncertainty is found if parameters obtained for $N \leqslant 126$ nuclei are used to describe the half-lives of $N > 126$ nuclei. Based on the Bohr-Sommerfeld quantization condition and simple assumptions, a detailed shell correction term is obtained for $N > 126$ nuclei, of which the value is similar to that in NGNL. A global hindrance on the $\alpha$ decay process is found in oA\_nc (oA\_c) nuclei comparing with ee (oA\_nc) nuclei. If parameters obtained from ee (oA\_nc) nuclei are used, the half-lives of oA\_nc (oA\_c) nuclei are generally underestimated with large systematic uncertainties, which can be related to the contribution of pairing effect and angular momentum. The parameter of angular momentum term in NGNL is obtained with large uncertainty and very sensitive to the selections of the dataset. The recently observed superallowed decay from $^{104}$Te to $^{100}$Sn is also discussed based on uncertainty analysis.
\item[Conclusions] The theoretical uncertainty of existing $\alpha$ decay formulas is successfully evaluated by the non-parametric bootstrap method, which simultaneously indicates the important effect in $\alpha$ decay, such as the shell effect and the pairing effect. In addition, statistical results show strong correlations between the parameters of the second and third term in both UDL and NGNL, which demands further investigations.
\end{description}
\end{abstract}

\pacs{}

\maketitle



%
%
%
%
\section{Introduction}
\label{sec01}

	Since the famous Geiger-Nuttall law was published in the 1910s, $\alpha$-decay half-life has been studied by physicists for more than a century. To extend the linear and simple empirical relationship to a larger range of nuclei, many effective generalizations, such as the Viola-Seaborg formula \cite{viola1966university}, were proposed. Recently, Poenaru \emph{et al.} introduced the Semi-Fission formula and the Universal curve for $\alpha$ decay based on fission theory \cite{poenaru2006alpha}. Qi \emph{et al.} deduced a Universal Decay Law (UDL) from R-matrix theory \cite{qi2009universal, qi2009microscopic}. Roger proposed an \emph{l}-dependent analytic formula \cite{royer2010analytic}, and Ren \emph{et al.} correct the original Geiger-Nuttall law with the preformation factor \cite{ni2008unified} as well as some important quantum numbers \cite{ren2012new}. All these formulas are in the form of linear combinations with several coefficients. Therefore, some parameters remain to be determined via a fitting process, on which the predictive ability and the robustness of such a semi-empirical or phenomenological model depend. As is shown correspondingly in different references, the results of all the mentioned formulas are fairly well.

	As the experimental data accumulates, the model validity would be tested by comparing a calculated value to an experimental result. Interestingly, one always tends to put a single, determined calculated value aside from an experimental error bar in practice. This act could lead to a confusing conclusion in some subtle cases \cite{qi2019recent}. All the justification should be accompanied with confidence from the statistical perspective. Hence, it is of great importance to implement a further investigation in detail on the uncertainty of a theoretical model. Our purposes are the following:
	
	1. to estimate the confidence interval of the predicted $\alpha$ decay half-lives;
	
	2. to identify the sources of uncertainty for a given formula;
	
	3. to figure out reliability, prosperity, and deficiency of existing models and possible improvements
	
	Some pioneering works in other fields have shown ways to study the uncertainty with traditional statistics tools including sensitivity analysis and Bayesian methods \cite{dobaczewski2014error,mcdonnell2015uncertainty,gao2013propagation,furnstahl2015recipe,schunck2015error,yuan2016uncertainty}. Besides these widely used tools, one can use the non-parametric bootstrap method on uncertainty analysis. As a reasonably accurate estimator widely accepted in the validation for machine learning strategies \cite{bisani2004bootstrap, mazucheli2005bootstrap, tiwari2010uncertainty, kybic2010bootstrap, zio2006study}, the Bootstrap method can simultaneously estimate statistical bias and reproduce the distribution of several given observables without any specific prior assumptions. The method has been tested in an analysis of NN scattering data by P{\'e}rez \emph{et al.} \cite{perez2014bootstrapping} and in studying nuclear mass models \cite{qi2019add}. In this paper, we further investigate its ability to explore the parameter space and to extract the uncertainty information.
	
	With present computational power, One can easily implement a powerful non-parametric bootstrap framework. Details of it will be presented in Sec. III. For the sack of simplicity, the authors have focused the study on UDL and the New Geiger-Nuttall Law (NGNL). A brief introduction is given in Sec. II. Both two laws would be tested with a dataset containing 162 even-even $\alpha$-emitters  in Sec. \ref{sec04} and 92 $\alpha$-decay channels of odd-mass nuclei in Sec. V. Physical essences revealed in uncertainty analysis would also be discussed. In the end, a summary of this article is given in Sec. VI.

%
%
%
%
%
\section{A brief introduction on dataset and models}
\label{sec02}

	The present work focuses on the uncertainty analysis of the $\alpha$-decay half-life estimation. Among the various models, UDL and NGNL are selected for their simplicity and clarity.
	
	Deduced from the R-matrix theory, UDL is read as following \cite{qi2009universal,qi2009microscopic}:
	\begin{equation} \label{udl}
	\log{T_{1/2}} = a Z_{c} Z_{d} \sqrt{\frac{A}{Q_{\alpha}}}+b \sqrt{A Z_{c} Z_{d} \left(A_{c}^{1 / 3}+A_{d}^{1 / 3} \right) } + c,
	\end{equation}
while NGNL taking the form \cite{ren2012new}:
	\begin{equation} \label{ngnl}
	\log T_{1 / 2}=a Z_{c} Z_{d} \sqrt{\frac{A}{Q_{\alpha}}}+b \sqrt{A Z_{c} Z_{d}}+c+S+P l(l+1).	
	\end{equation}
	
	In both equations, $\displaystyle A = \frac{A_{c}A_{d}}{A_{c} + A_{d}}$ denotes the reduced mass number, where the subscripts $c$ and $d$ correspond \emph{cluster} and \emph{daughter} nuclei, respectively. $a$, $b$, $c$ are undetermined coefficients in which one interests. Among the two, NGNL is more complicated. The term $S$ here deals with the shell effect, where $S=0$ for nuclei with $\text{N} > 126$ and $S=1$ for $\text{N} \leqslant 126$. $l$ is the quantized angular momentum, and $P$ is one more undetermined coefficient related to the parity. In the case of the $\alpha$ decay, one gets $Z_{c}=2$ and $A_{d}=4$.
	
	Before further discussion, physical significance should be explained for each term. Indeed, both of the two laws are generalizations of the original Geiger-Nuttall law. Thus, the first term could be regarded as a description of the quantum tunneling process, as is demonstrated by Gamow in 1928, while the second term involving the formation probability. The coefficient $a$, $b$, $c$ describe some global average properties and could be written in the explicit form (see Ref.~\cite{qi2009microscopic} for UDL and Ref.~\cite{ni2008unified} for NGNL).
	
	As for the dataset taken in regression, NUBASE2016 \cite{audi2017nubase2016} and AME2016 \cite{huang2017ame2016,wang2017ame2016} provided by Atomic Mass Data Center are selected for the analysis of even-even (ee) nuclei. For better validation, all the undetermined experimental results (marked with "*" in the dataset), as well as extrapolation results (marked with "\#" in the dataset), are removed. As a result, a total of 162 ee nuclei are included for further analysis. The case of odd-mass (oA) nuclei is more complicated. Since the competition between the decay to ground state and to excited state plays an important role here, the present work takes the different decay channels into account. All those data are collected from National Nuclear Data Canter. Similar filtration is performed with an extra rule: only channels with branch ratio greater than 1\% are reserved. Finally, the database we use for analysis consists of 162 ee nuclei decaying from ground state to ground state (Gs-Gs), 42 channels of ee nuclei of which $N>126$ and $l=2$ decaying from ground state to excited state (Gs-Ex) and 92 channels of 46 odd-mass (oA) nuclei. A review of several special outliers will be given in the following discussions.

%
%
%
%
\section{Non-parametric bootstrap and uncertainty decomposition}
\label{sec03}

	We aim to estimate the uncertainty of a given parametric model with several parameters waiting to be determined via fitting. Such a theoretical model are in the form of:
	\begin{equation}
		y = y\left(\vec{x};\vec{p}\right)+u,
	\end{equation}
where $\vec{x}$ denotes the input values and $\vec{p}$ the parameter. In the present work, the input values are the nuclear properties and the parameters are the corresponding coefficients. $u$ denotes the unspecified deficiency, in other words, the total uncertainty. In general, it is a random variable under certain distribution.

	In this case, the total uncertainty of a predictive result is known as a mixture of three parts. The first part is systematic, which is due to the deficiency of the model. The second one is statistical, which originates from undetermined parameters. The last part is the experimental uncertainty. For the present dataset, the variations of the measured decay energy are very small in general. The variations of $T_{1/2}$ of several nuclei ($^{178}$Pb, $^{228}$Pu, etc.) are, on the other hand, comparable to those of the theoretical ones. For simplicity, all data are regarded as exact values in the most analyses presented in this work. Meanwhile, test analysis is also implemented based on the present dataset to examine the uncertainty induced by experimental uncertainties.
	\begin{figure*}
		\includegraphics[width=.95\textwidth]{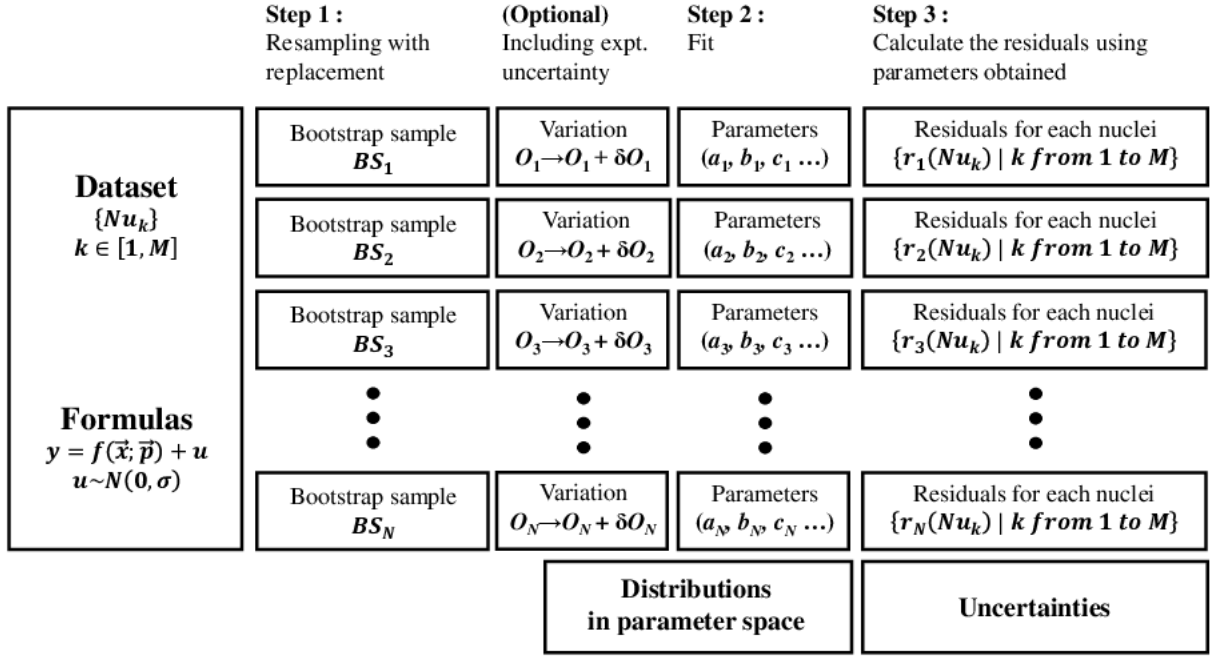}
		\caption{\label{3-step-bootstrap}An illustration of the 3-step bootstrap method}
	\end{figure*}

	Traditional statistical methods such as the linear fitting provide tools to access the variation of parameters and bias between theoretical and experimental results. The key technique here, the resampling, further reconstructs the parameter space and records simultaneously the uncertainty information. It can be described as a 3-step calculation like the following:
	
	\textbf{Step 1.} From a given dataset, e.g. the 162-term ee nuclei dataset, one can always extract N samples with the same sample size 162, by resampling with replacement (a data can be selected more than once during resampling). From a statistical point of view, it is reasonable to keep the size and the degrees of freedom of each bootstrap sample the same as those of the original dataset. The statistical result is obviously more precise when N takes larger values.

As verification, available experimental uncertainties are also taken into consideration. With the normality assumption, additional variations are introduced to every observable (decay energy and half-life) used in each bootstrap sample,
	\begin{equation}
	\begin{aligned}
		\mathcal{O}(Nu_{k}) \to \mathcal{O}(Nu_{k}) + \delta \mathcal{O}(Nu_{k}),     \\
		\delta \mathcal{O}(Nu_{k}) \sim \mathcal{N}(0,\sigma_{\text{exp}}(Nu_{k})),
	\end{aligned}
	\end{equation}
where $\sigma_{\text{exp}}(Nu_{k})$ denotes the corresponding experimental uncertainty. It should be emphasized that the variation takes different values for different observables in different bootstrap samples, to take into account the random effect induced by experimental uncertainty. In the present study, the experimental uncertainty of decay energy $E$ and/or of half-life $T_{1/2}$ is included, while that of intensity $I$ is not due to the lack of information of its uncertainty in the data library. Now the preparation is completed. The extracted samples named bootstrap sample by Efron \cite{efron1982jackknife} would play the role of a test set in the following calculation.

	\textbf{Step 2.} A fitting process is implemented on each bootstrap sample with all results, i.e. fitting parameters, saved. Here distributions on parameter space are approached by counting the parameter arrays obtained.
	
	\textbf{Step 3.} Hence, with the obtained parameters, it is easy to compute predictive half-lives for all nuclei in the original dataset. Comparing with the experimental values, residuals are derived. All information about the uncertainty is in fact encoded in it.

	Intuitively, resampling is comparable to experimental measurement. While measuring an unknown physical quantity, a natural strategy is repeating several identical measurements upon the same object. Likewise, the Bootstrap method could be considered as a Monte-Carlo event generator. While repeating resampling and calculation described above, it is like to repeat virtual measurements upon the given theoretical model. With sufficient replications, one can safely approximate the distribution of any observables. For summarizing, a flow chart is presented in Fig. \ref{3-step-bootstrap}. In practice, the replication quantity is taken to be $N=10^{6}$ with a computation facility.
	
	In Step 3, the residual between the recorded experimental value and the measured values is defined as:
	\begin{equation}
		r\left(N u_{k}, B S_{i}\right)=\log T_{\mathrm{exp}}\left(N u_{k}\right)-\log T_{\text{cal}}\left(N u_{k}, B S_{i}\right),
	\label{def_of_res}
	\end{equation}
where $Nu_{k}$ indicates the $k$-th nucleus in the original dataset and $BS_{i}$ denotes different bootstrap sample with $i$ ranging from 1 to $N$. Our goal is to extract the different types of uncertainties from cards in hand. Naturally, the estimated value is set as the center of all predicted value $\bar{r}\left(N u_{k}\right) = \sum_{i=1}^{N} r\left(N u_{k}, B S_{i}\right) / N$ and the statistical uncertainty indicates the dispersion of results. In other words, it is the unbiased standard deviation:
	\begin{equation}
		\hat{\sigma}_{\operatorname{stat}}^{2}\left(N u_{k}\right)=\frac{1}{N-1} \sum_{i=1}^{N}\left(r\left(N u_{k}, B S_{i}\right) - \bar{r}\left(N u_{k}\right)\right)^{2}.
	\label{sigma_stat_single}
	\end{equation}
On the other hand, the systematic uncertainty yields gaps between the estimated value and the ``true'' value. After simple calculation, one might find that it is equal to the absolute value of the mean of residuals we concerned:

	\begin{equation}
		\begin{aligned}
		\hat{\sigma}^{2}_{sys}\left(N u_{k}\right)&=\left(\sum_{i=1}^{N} \frac{\log T_{\text{cal}}\left(N u_{k}, B S_{i}\right)}{N}-\log T_{\exp }\left(N u_{k}\right)\right) ^{2}	\\
		&= \left( \frac{1}{N} \sum_{i=1}^{N} r\left(N u_{k}, B S_{i}\right) \right)^{2}.
		\end{aligned}
	\label{sigma_sys_single}
	\end{equation}

	Noticed that until now, what we obtained is the uncertainty for a specific nucleus. However, the systematic uncertainty is more important for a whole dataset rather than a single data. The formulas just derived is merely an intermediate result and let us focus on the following definition:
	\begin{equation}
		\hat{\sigma}_{sys}^{2}=\frac{1}{M} \sum_{k=1}^{M} \hat{\sigma}^{2}_{sys}\left(N u_{k}\right) = \frac{1}{M} \sum_{k=1}^{M} \bar{r}^{2}\left(N u_{k}\right)
	\label{sigma_sys_all}
	\end{equation}
The coefficient $1/M$ emphasizes the fact that every nucleus contributes equally to resampling. In the case of Ordinary Least Square fitting, such a definition is asymptotically coincident to the RMS metric when quantity $N$ goes to infinity.

Total uncertainty for a single nucleus is evaluated as:
	\begin{equation}
		\begin{aligned}
		\sigma_{\text {total}}^{2}\left(N u_{k}\right)  & =\frac{1}{N} \sum_{i=1}^{N} r^{2}\left(N u_{k}, B S_{i}\right) \\
		                                                                    & =\frac{N-1}{N} \hat{\sigma}_{stat}^{2}\left(N u_{k}\right)+\hat{\sigma}_{sys}^{2}\left(N u_{k}\right).
		\end{aligned}
	\label{sigma_total_all}
	\end{equation}
Similarly, it can be generalized to the whole dataset, as well as the statistical uncertainty. Thus it is easy to verify the following relation holds in a global sense:
	\begin{equation}
		\hat{\sigma}_{\text{total}}^{2} = \hat{\sigma}_{\text{sys}}^{2} + \hat{\sigma}_{\text{stat}}^{2}.
	\label{sigma_relationship}
	\end{equation}

	Besides the uncertainty evaluation for the theoretical descriptions on the known properties, another important task is to assign confidence interval to each predicted result, which should be a composition of the global systematic bias due to model deficiency and the statistical uncertainty assigned to the specific nucleus. It is considered to take the following form (the approximation is due to the large resampling quantity):
	\begin{equation}
		\hat{\sigma}_{pred}^{2}\left(N u_{k}\right) \approx \hat{\sigma}_{s t a t}^{2}\left(N u_{k}\right)+\hat{\sigma}_{s y s}^{2}.
	\label{sigma_pred}
	\end{equation}
Now one can apply the framework to study the predictive power of UDL and NGNL.

%
%
%
%

\begin{table*}[]
\caption{Parameters obtained from different even-even nuclei datasets for both UDL and NGNL and corresponding uncertainties.}
\label{param_digit_ee}
	\begin{threeparttable}
		\begin{ruledtabular}
			\begin{tabular}{cccccccccccccc}
			 &  & \multicolumn{2}{c}{a} & \multicolumn{2}{c}{b} & \multicolumn{2}{c}{c} & \multicolumn{2}{c}{d} & Residuals& \multicolumn{3}{c}{Uncertainty} \\
			 &  & Mean & S.D. & Mean & S.D. & Mean & S.D. & Mean & S.D. & Mean& total & stat & sys \\
			\colrule
			\multirow{8}{*}{UDL} & (Gs-Gs) & 0.4096 & 0.0022 & -0.4266 & 0.0043 & -21.5675 & 0.2880 &\textemdash &\textemdash &\textemdash & 0.3442 & 0.0503 & 0.3405 \\
			& (Gs-Gs)\tnote{T} & 0.4096 & 0.0022 & -0.4266 & 0.0044 & -21.5766 & 0.2955 &\textemdash &\textemdash &\textemdash  & 0.3444 & 0.0513 & 0.3406 \\
			& (Gs-Gs)\tnote{E} & 0.4096 & 0.0022 & -0.4265 & 0.0044 & -21.5665 & 0.2910 &\textemdash  &\textemdash&\textemdash  & 0.3443 & 0.0506 & 0.3405 \\
			& (Gs-Gs)\tnote{1} & 0.4101 & 0.0022 & -0.4270 & 0.0043 & -21.6044 & 0.2914 &\textemdash  &\textemdash&\textemdash  & 0.3479 & 0.0506 & 0.3442 \\
			 & $N\leqslant126$ & 0.4188 & 0.0031 & -0.3943 & 0.0054 & -24.7439 & 0.5106 &\textemdash  &\textemdash&\textemdash  & 0.2922 & 0.0626 & 0.2854 \\
			  & $N\leqslant126\tnote{1}$ & 0.4188 & 0.0031 & -0.3943 & 0.0054 & -24.7426 & 0.5073 &\textemdash  &\textemdash&\textemdash  & 0.2915 & 0.0631 & 0.2846 \\		
			 & $N> 126$ & 0.4013 & 0.0013 & -0.3707 & 0.0066 & -24.7370 & 0.5221 &\textemdash  &\textemdash&\textemdash  & 0.1737 & 0.0356 & 0.1700 \\
			 & $N> 126$\tnote{1} & 0.4041 & 0.0027 & -0.3897 & 0.0180 & -23.6893 & 1.0640 &\textemdash  &\textemdash&\textemdash  & 0.2424 & 0.0686 & 0.2325 \\
			 & $N> 126$\tnote{*} &\textemdash  &\textemdash  &\textemdash  &\textemdash  &\textemdash  &\textemdash  &\textemdash  &\textemdash &-0.6619 & 0.7330 & 0.0939 & 0.7270 \\
			\colrule
			\multirow{8}{*}{NGNL} & (Gs-Gs) & 0.4075 & 0.0019 & -1.3255 & 0.0125 & -17.7229 & 0.2150 &\textemdash &\textemdash &\textemdash & 0.2658 & 0.0407 & 0.2626 \\
			& (Gs-Gs)\tnote{T} & 0.4076 & 0.0020 & -1.3256 & 0.0131 & -17.7309 & 0.2301 &\textemdash &\textemdash &\textemdash & 0.2661 & 0.0421 & 0.2627 \\
			& (Gs-Gs)\tnote{E} & 0.4075 & 0.0019 & -1.3255 & 0.0127 & -17.7223 & 0.2214 &\textemdash &\textemdash &\textemdash & 0.2658 & 0.0410 & 0.2626 \\
			& (Gs-Gs)\tnote{1} & 0.4083 & 0.0020 & -1.3280 & 0.0125 & -17.7649 & 0.2223 &\textemdash &\textemdash &\textemdash & 0.2807 & 0.0428 & 0.2774 \\
			 & $N\leqslant126$ & 0.4126 & 0.0030 & -1.3574 & 0.0143 & -17.6370 & 0.4536 &\textemdash &\textemdash &\textemdash & 0.2736 & 0.0577 & 0.2674 \\
			 & $N\leqslant126\tnote{1}$ & 0.4126 & 0.0029 & -1.3574 & 0.0143 & -17.6363 & 0.4515 &\textemdash &\textemdash &\textemdash & 0.2724 & 0.0582 & 0.2661 \\
			 & $N> 126$ & 0.3982 & 0.0013 & -1.2622 & 0.0213 & -18.0730 & 0.5863 &\textemdash &\textemdash &\textemdash & 0.1720 & 0.0344 & 0.1686 \\
			 & $N> 126$\tnote{1} & 0.4008 & 0.0026 & -1.3251 & 0.0620 & -16.7341 & 1.3852 &\textemdash &\textemdash &\textemdash & 0.2454 & 0.0694 & 0.2354 \\
			 & $N> 126$\tnote{*} &\textemdash &\textemdash &\textemdash &\textemdash &\textemdash &\textemdash &\textemdash &\textemdash &0.1093 & 0.3079 & 0.0814 & 0.2970 \\
			\colrule
	uncorrected (Eq. \ref{uncorrected})& (Gs-Gs) & 0.4031 & 0.0024 & -1.5054 & 0.0190 & -11.9837 & 0.4932 &  \textemdash  &  \textemdash&  \textemdash  & 0.3939 & 0.0594 & 0.3894	\\
	corrected   (Eq. \ref{corrected})& (Gs-Gs) & 0.4069 & 0.0020 & -1.3682 & 0.0142 & -16.4383 & 0.3919 & 1.7018 & 0.1278 &  \textemdash & 0.2657 & 0.0466 & 0.2616
			\end{tabular}%
		\end{ruledtabular}
		\begin{tablenotes}
			\footnotesize
			\item[*] Uncertainty estimated by parameters obtained from set $N\leqslant126$ as an extrapolation.
			\item[T] Bootstrap with consideration of half-life experimental uncertainty.
			\item[E] Bootstrap with consideration of $\alpha$ decay energy experimental uncertainty.
			\item[1] Dataset added $^{210}$Pb.

		\end{tablenotes}
	\end{threeparttable}
\end{table*}

\section{Application to even-even nuclei}
\label{sec04}

	In this section, the established non-parametric bootstrap method is tested by applying it to both UDL and NGNL with a dataset consisting of 162 ee nuclei. Firstly, the resampling is applied to the whole dataset. Then, the shell effect is accounted to investigate its impact in decay half-life prediction. Since we noticed that the neutron shell plays a more important role than that of proton in $\alpha$ decay theory \cite{buck1991ground}, the original dataset is divided into two parts according to the neutron number, (i) the set for ee nuclei with $\text{N} \leqslant 126$  and (ii) the set for ee nuclei with $\text{N} > 126$. The same resampling method will be applied respectively to these two subsets. Furthermore, another dataset composed of 42 $\alpha$-decay channels of ee nuclei with $N>126$ and $l=2$ is investigated and compared with the results from the ground state to ground state.

	\subsection{Exploration in parameter space and uncertainty}
	
	\begin{figure*}
		\centering
		\includegraphics[width=.31\textwidth]{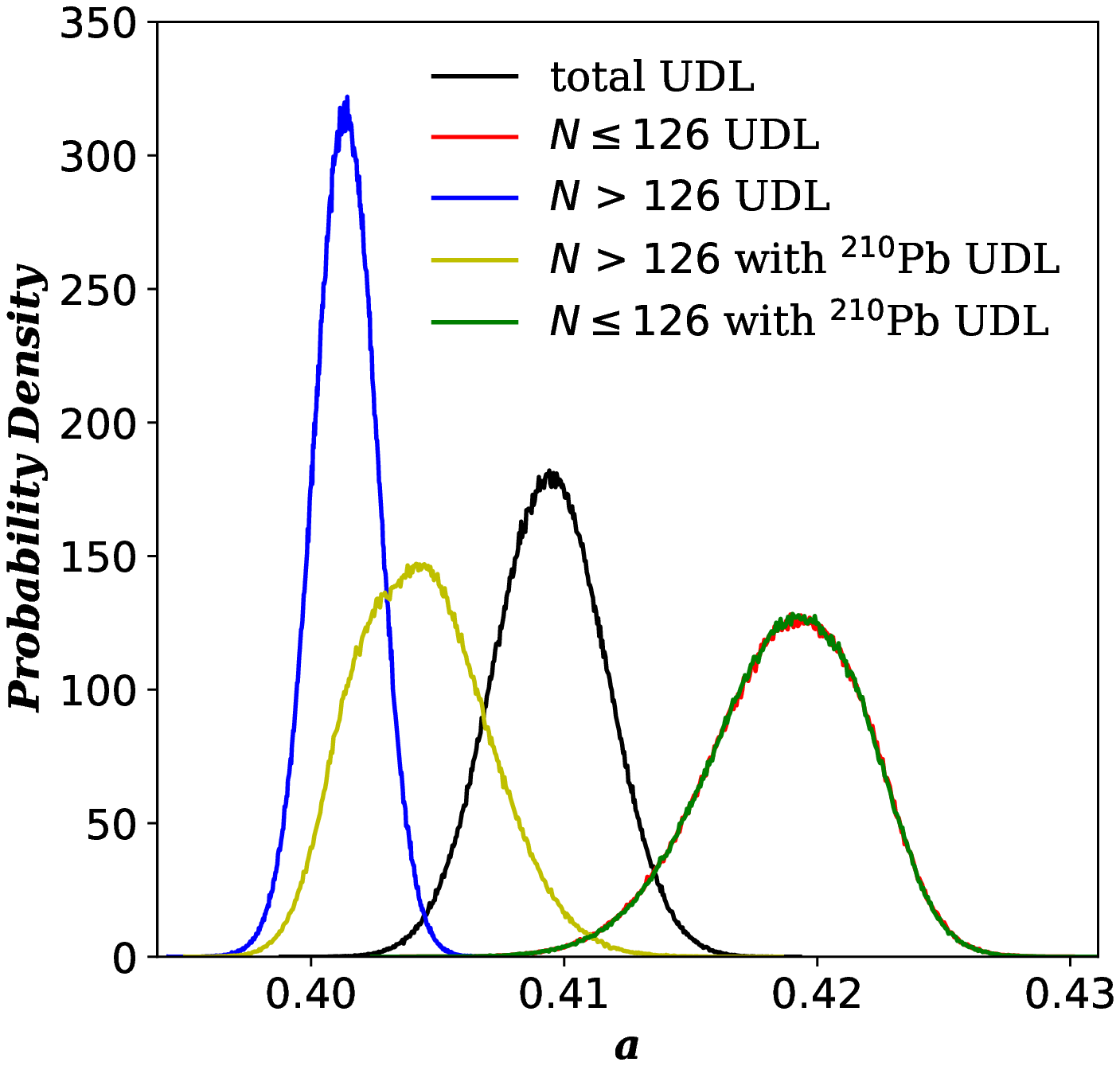}
		\includegraphics[width=.31\textwidth]{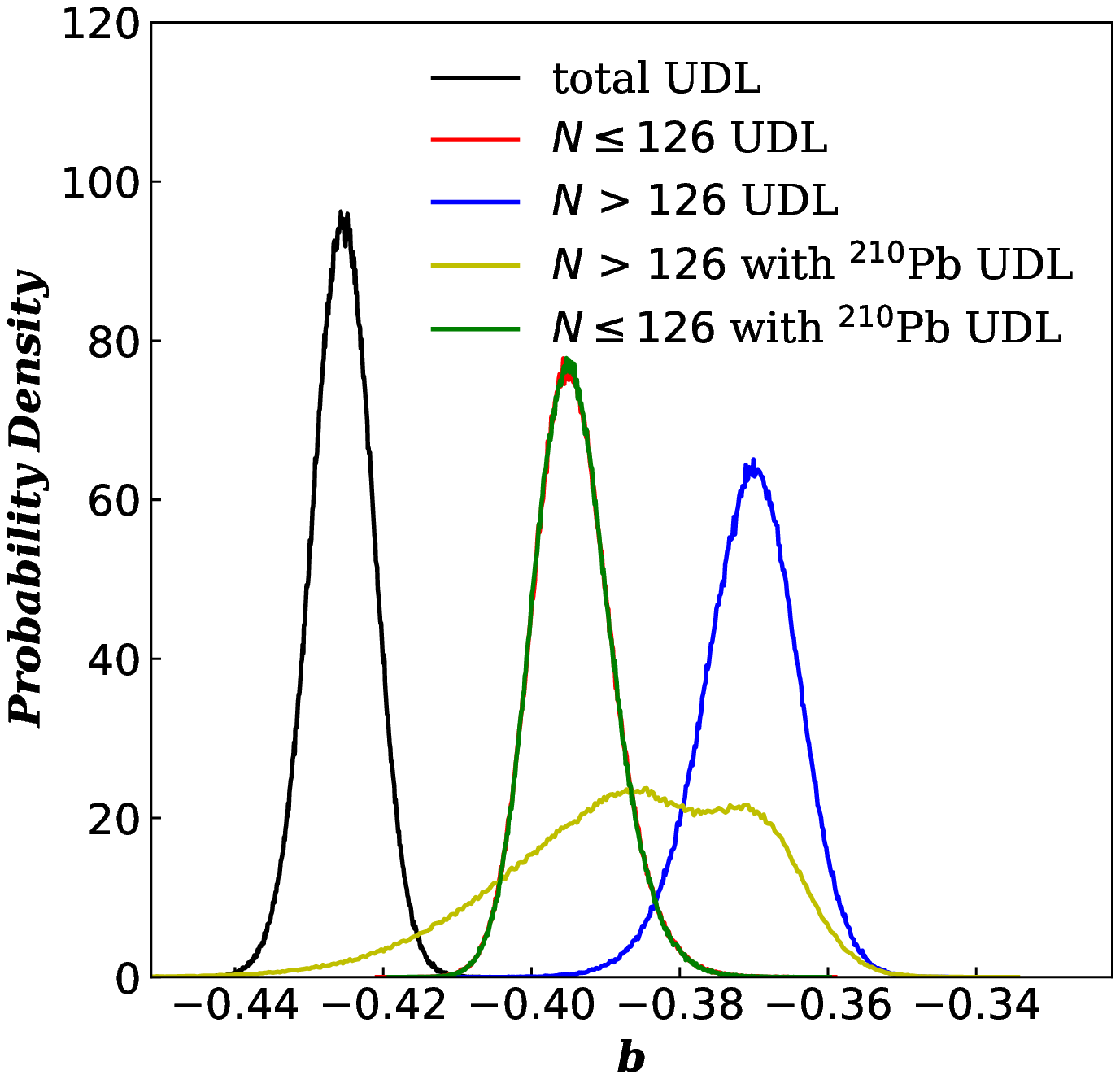}
		\includegraphics[width=.31\textwidth]{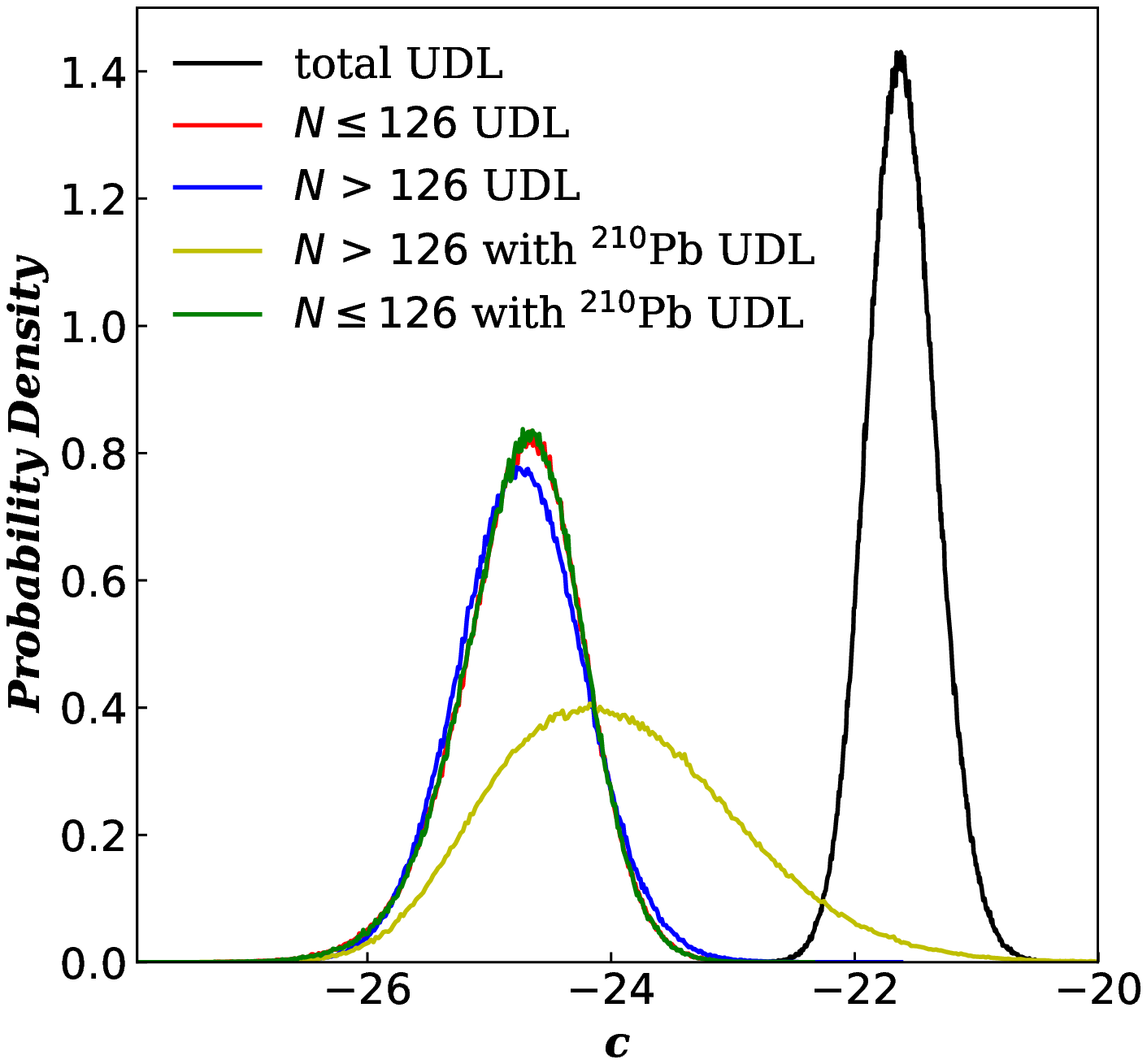} \\	
		\includegraphics[width=.31\textwidth]{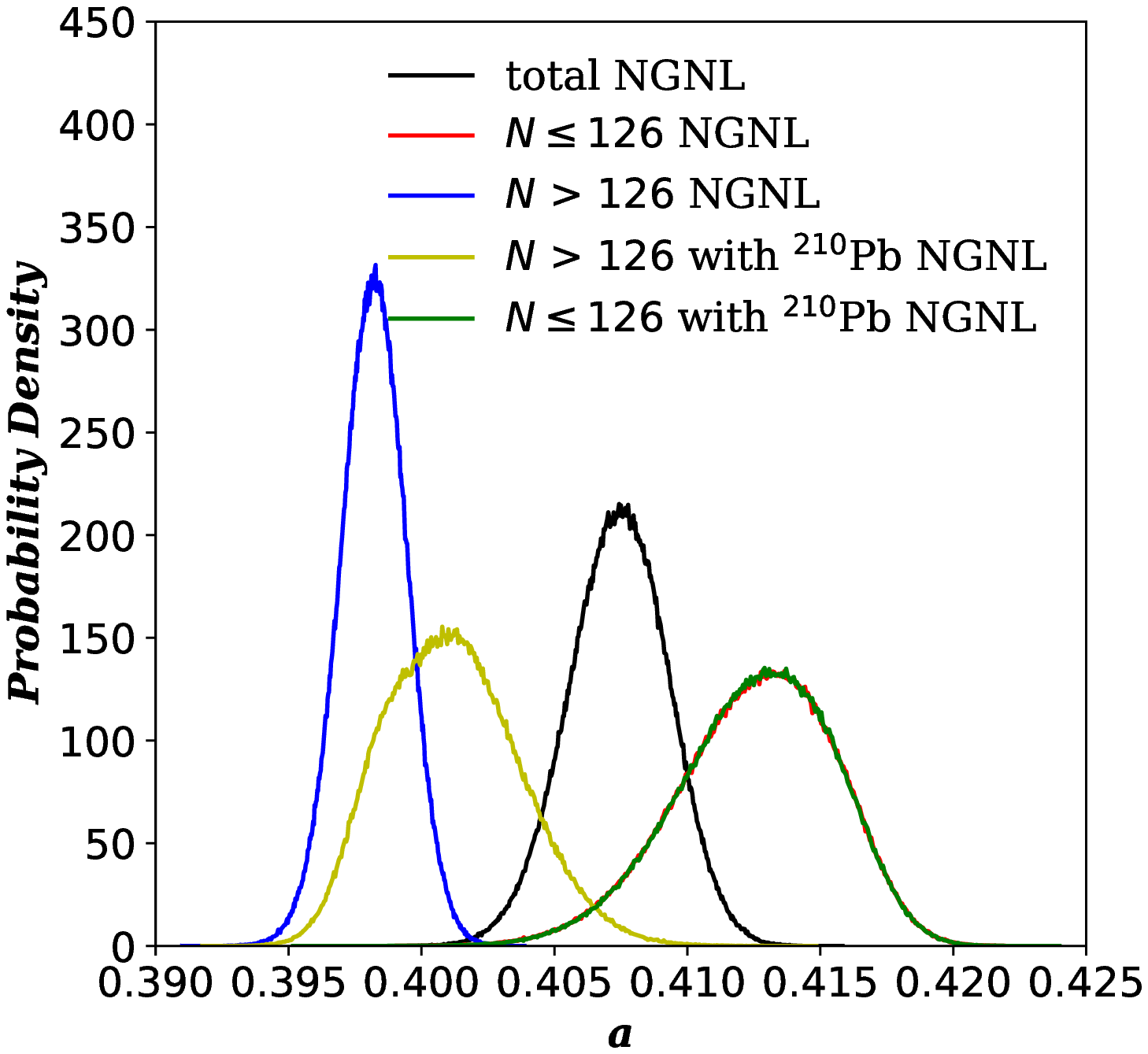}
		\includegraphics[width=.31\textwidth]{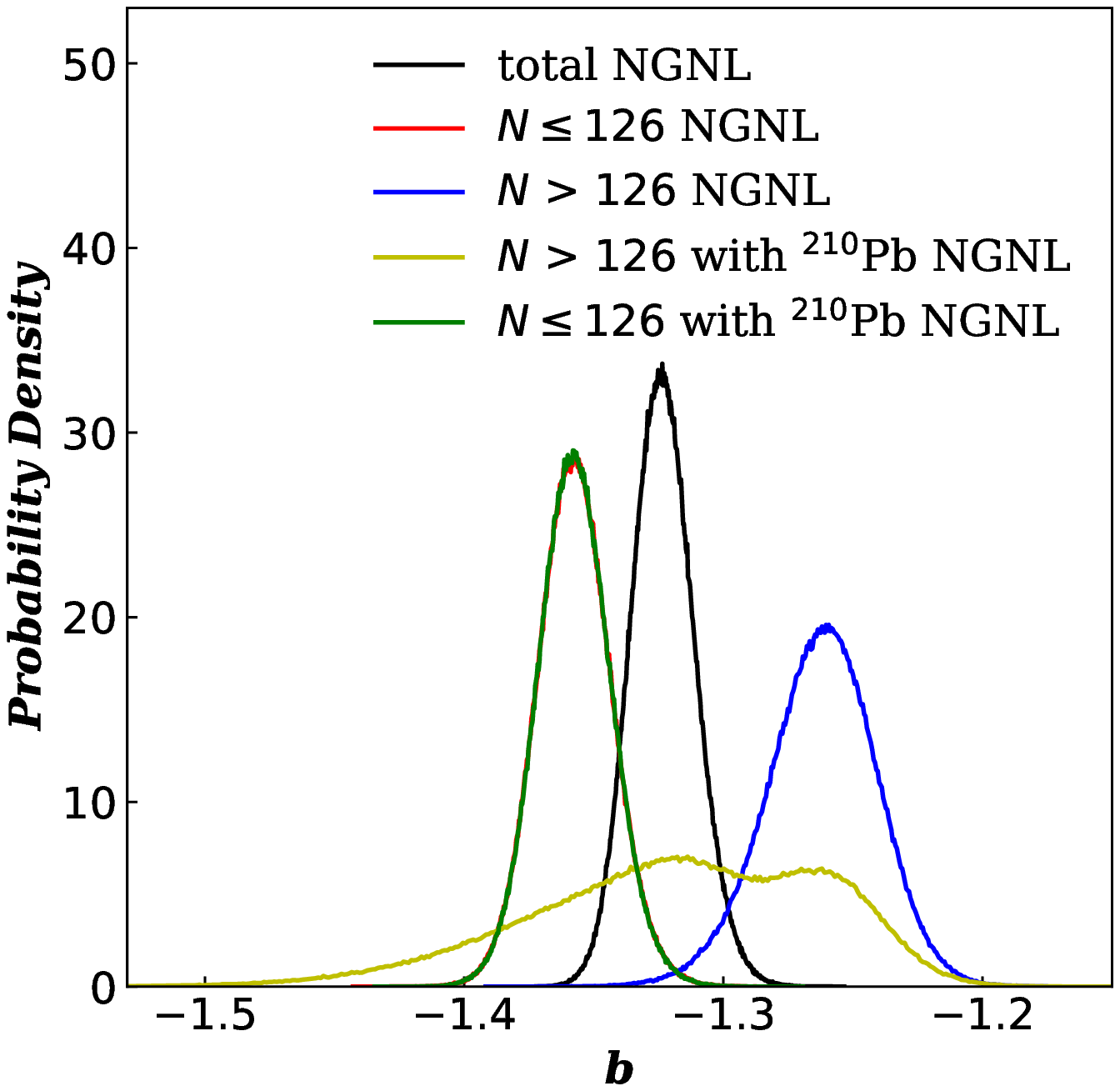}
		\includegraphics[width=.31\textwidth]{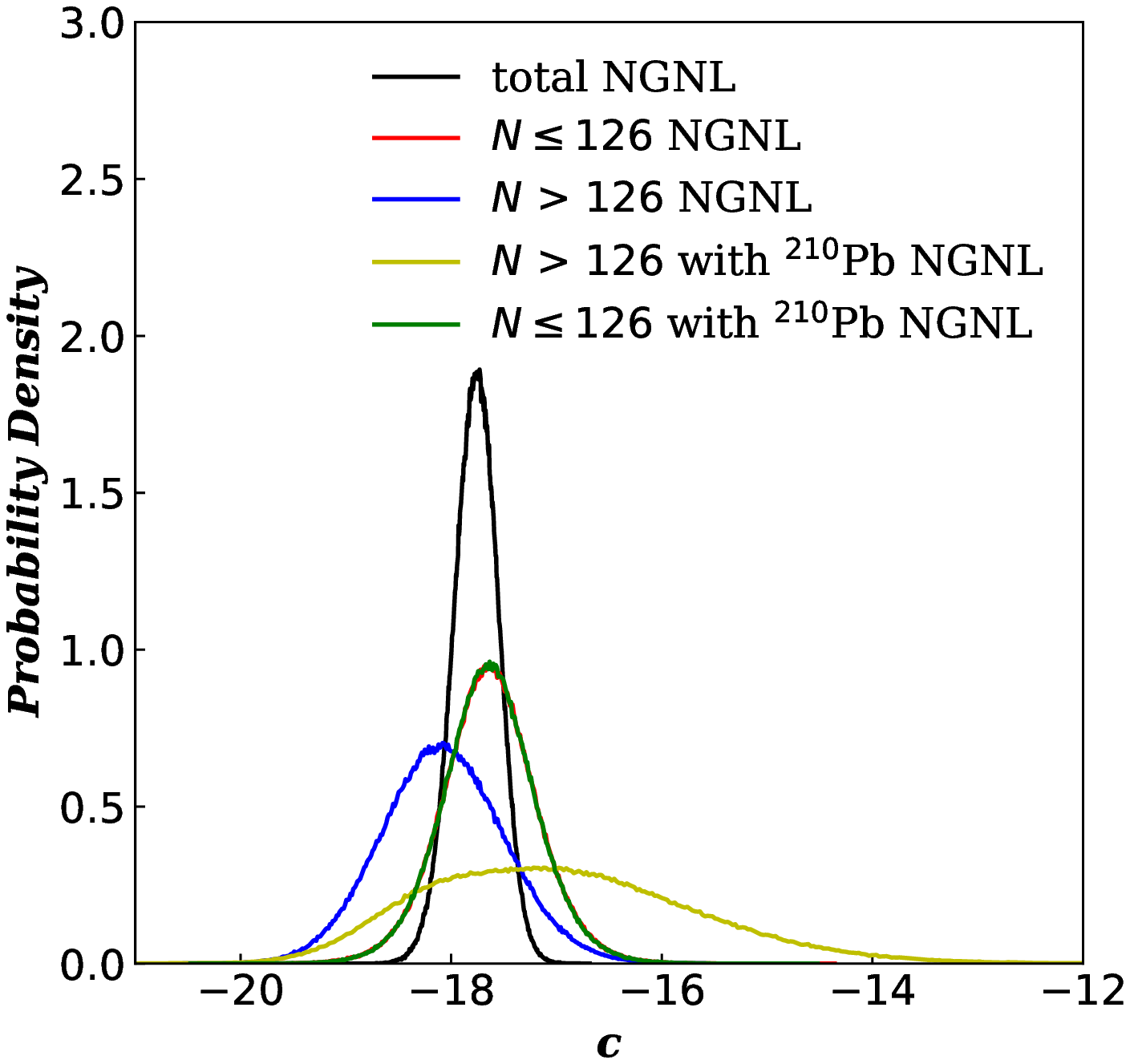}\\
		\caption{\label{param_dist_ee}Distributions of parameters for UDL (top) and NGNL (bottom), fitted to even-even nuclei. From left to right, three figures invoke parameter $a$, $b$, $c$ in turn. In figures, black line for all nuclei, red line for nuclei with $\text{N} \leqslant 126$, blue line for nuclei with $\text{N} > 126$ and yellow line for nuclei with $\text{N} > 126$ including $^{210}$Pb.}
	\end{figure*}	
	
	Fig. \ref{param_dist_ee} shows the reconstructed marginal distribution of each parameter in the corresponding parameter space for both UDL and NGNL. Curves except the $\text{N} > 126$ results with $^{210}$Pb approximate the normal distributions, which indicates that both UDL and NGNL sketched out common decay features with stochastically extended parameters. The results also support the normality assumption in traditional statistical methods. As for the dataset $\text{N} > 126$ with $^{210}$Pb, its distribution with large extension indicates that $^{210}$Pb is an outlier in this dataset. It is the only nuclide with $\text{Z}=82$ and $\text{N} > 126$, whose decay feature is slightly different from other nuclides in the same set. Here the present framework works as an indicator of outliers. The expected distribution can still be extracted from the deformed case, because the resampling strategy takes a sample without $^{210}$Pb with a probability $\displaystyle P=\left( 1 - \frac{1}{M} \right) ^{M} \approx e^{-1}$. The distribution for the dataset $\text{N} > 126$ with $^{210}$Pb in Fig. \ref{param_dist_ee} can be considered as a combination of two peaks, from the bootstrap samples with and without the outlier. One can see that the additional $^{210}$Pb tends to incline the regression plane further, as is discussed in Ref.~\cite{qi2014validity}.

	Traditional statistics, including the mean values and the standard deviations (S.D.), are taken to sketch out the approximately normal distributions. Otherwise, one should calculate the percentile via numerical methods to determine the confidence interval in the case of the deformed distribution. The mean values and S.D. of each parameter in UDL and NGNL are summarized in TABLE \ref{param_digit_ee}. The comparison of the results shown in the first three rows confirms the assumption that the experimental uncertainty can be negligible in our analysis. Moreover, all parameters take relatively small S.D. comparing with their mean values. The constant term takes the largest uncertainty, which suggests that the preformation of $\alpha$ clusters is not well described and some properties of nuclei may need to be included, such as the deformation. It is clearly seen that both UDL and NGNL result in rather a similar parameter $a$ but quite different $b$ and $c$, which comes from the selection of different radius models and will be discussed later.

If the "outlier" nucleus $^{210}$Pb is added to all dataset and the dataset with  $\text{N} \leqslant 126$, the fitting results change little as seen in TABLE \ref{param_digit_ee}. But if $^{210}$Pb is added to the dataset with $\text{N} > 126$, the fitting results changes with a certain degree, as shown in both Fig. \ref{param_dist_ee} and TABLE \ref{param_digit_ee}. The comparison among these results indicates that the behavior of decay properties of $^{210}$Pb is not an outlier in all dataset but the dataset with $\text{N} > 126$. Hence, the decay property of $^{210}$Pb is more like those of nuclei with $\text{N} \leqslant 126$. This result is consistent with the microscopic calculations as presented in Refs. \cite{qi2019recent, qi2014validity}. It is related to the fact that the $\alpha$ formation probability is expected to be strongly suppressed at the $N=126$ and $Z=82$ shell closures (see for example Fig. 21 in Ref. \cite{qi2019recent}). In particular, the $\alpha$ formation probability in semi-magic nucleus $^{210}$Pb with $Z=82$ is expected to be similar to that of the nucleus $^{210}$Po with $N=126$ but is strongly suppressed in comparison to neighboring open-shell nuclei. Such suppression is due to the reduced pairing correlation caused by the large $Z=82$ proton shell gap in $Z=82$ and $N=126$ neutron shell gap in $^{210}$Po \cite{qi2010prc}.  In later discussion, $_{82}^{210}$Pb$_{128}$ is treated as a normal nucleus in the group with $\text{N} \leqslant 126$.

\begin{table}
\caption{Pearson matrix of parameters in UDL and NGNL, fitted with different datasets.}
\label{pearson_ee}
\begin{ruledtabular}
\begin{tabular}{ccccc|ccc}
                         &   & \multicolumn{3}{c}{UDL}     & \multicolumn{3}{c}{NGNL} \\
                         &   & a       & b       & c       & a           & b          & c    \\
\colrule
\multirow{3}{*}{(Gs-Gs)} & a & 1.000   &\textemdash &\textemdash & 1.000 &\textemdash &\textemdash \\
                         & b & -0.510  & 1.000   &\textemdash & -0.719   & 1.000      &\textemdash \\
                         & c & -0.460  & -0.525  & 1.000   & -0.107      & -0.611     & 1.000\\
\colrule
\multirow{3}{*}{$\text{N} > 126$}   & a & 1.000  &\textemdash &\textemdash & 1.000    &\textemdash  &\textemdash \\
                         & b & -0.123  & 1.000   &\textemdash & -0.163   & 1.000      &\textemdash \\
                         & c & -0.216  & -0.942  & 1.000   & -0.118      & -0.960     & 1.000\\
\colrule
\multirow{3}{*}{$\text{N} \leqslant 126$}   & a & 1.000   &\textemdash &\textemdash & 1.000           &\textemdash &\textemdash \\
                         & b & -0.084  & 1.000   &\textemdash & -0.200   & 1.000      &\textemdash \\
                         & c & -0.731  & -0.616  & 1.000   & -0.684      & -0.575     & 1.000\\
\end{tabular}
\end{ruledtabular}
\end{table}

	Another advantage of the present framework is to help investigate the correlation between pairs of parameters as an approximation of the given joint distribution easily. As an illustration, the calculated Pearson matrix in different cases is shown in TABLE \ref{pearson_ee}. All the parameters are mutually resistant. After classifying the nuclei into two different neutron shell, the correlation between parameters $a$ and $b$ is weakened, and, on the contrary, the correlation between parameters $b$ and $c$ is reinforced, upon which it is interesting to perform further investigation.

	As an important criterion, the decomposed uncertainties are also presented in TABLE \ref{param_digit_ee}. In general, NGNL gives a better description than UDL, according to a comparison of systematic uncertainties. Especially when the parameters of UDL are obtained from dataset $\text{N} \leqslant 126$ and used to describe the $\text{N} > 126$ data, the average value of the residuals is negative and far from zero. Thus a large systematic uncertainty is obtained, which indicates that UDL needs to include shell effect and will be discussed in the following section.

	On the contrary, NGNL shows its consistency among all results. Noticed that both two laws are more suitable for the heavier nuclei, a possible interpretation is that the $\alpha$ decay of heavy nuclei is dominated by the tunneling process. The existence of cluster structure facilitates the preformation part. Comparing with the systematic uncertainties, all statistical uncertainties are negligible, which indicates that the existing terms in both UDL and NGNL do reflect the physical nature of $\alpha$ decay of nuclides considered here. Such small statistical uncertainties ensure the stability of a global prediction on all nuclides.  But certain values of systematic uncertainties indicate that some more physical terms may need to be considered in the two laws.

	\subsection{Universality and shell correction}

	Since both laws are assumed to give a universal description at least for the even-even nuclei, it is of interest to investigate their performance while facing the shell closure. From Fig. \ref{param_dist_ee}, parameter $a$ and $b$ in both laws show certain deviations from $\text{N}> 126$ case to $\text{N}\leqslant 126$ case. If parameters obtained from the $\text{N}\leqslant 126$ data are used to describe the $\text{N}> 126$ data as an extrapolation shown in Table \ref{param_digit_ee}, the systematic uncertainty becomes much larger for UDL, while the uncertainty of NGNL is relatively small because of the shell correction term.

\begin{figure*}
	\includegraphics[width=.9\textwidth]{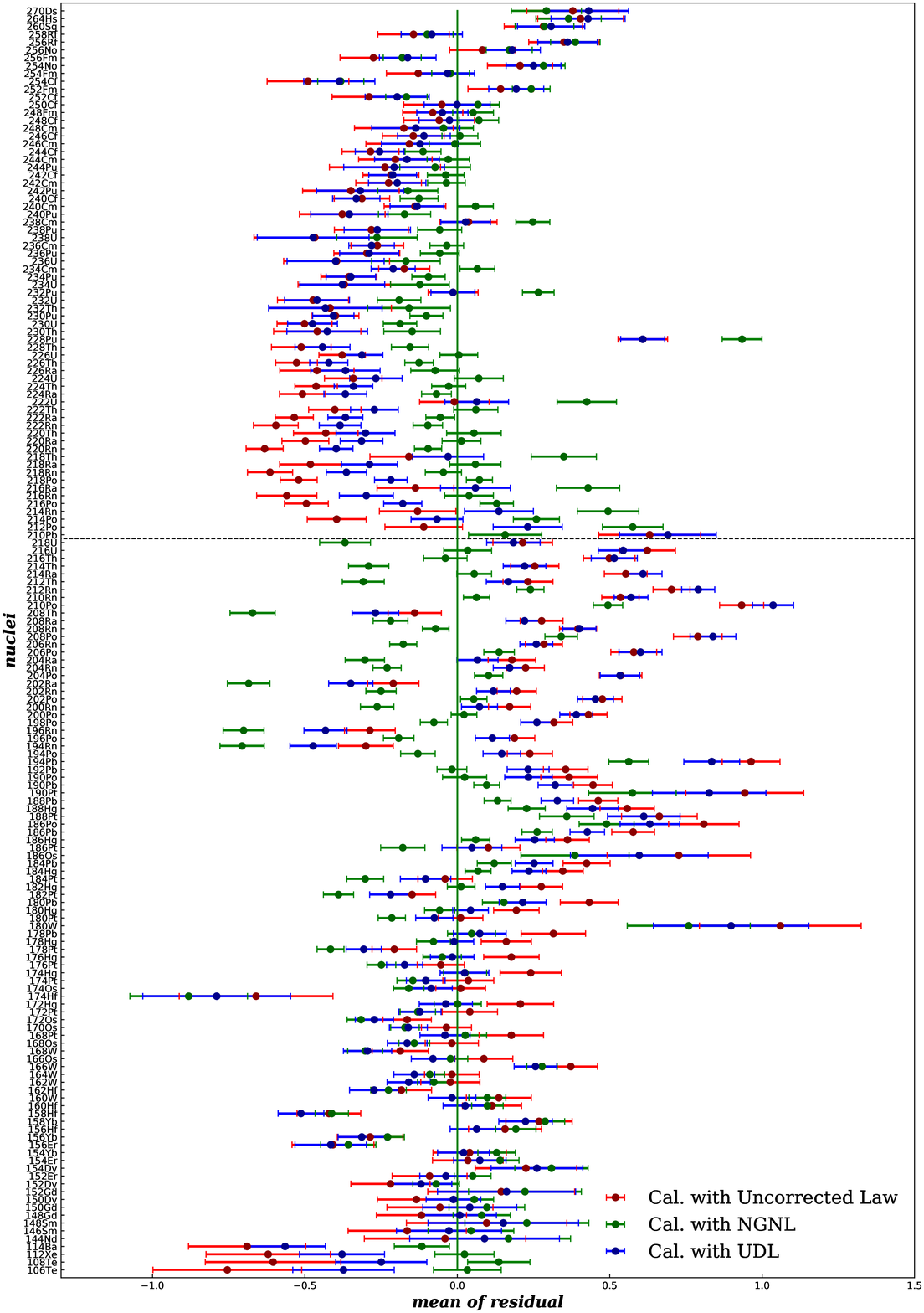}
	\caption{\label{3lawscomparison_ee}Comparison of predictive residuals for the uncorrected formulas (red), UDL (blue) and NGNL (green). The vertical dark green line indicates the coincidence of theoretical and experimental results. The black dash line in the middle denotes the shell closure $\text{N} = 126$.}
\end{figure*}
		
Removing the shell correction term S from NGNL, an uncorrected formula read as:
		\begin{equation}	
		\label{uncorrected}
		\log T_{1 / 2}=a Z_{c} Z_{d} \sqrt{\frac{A}{Q_{\alpha}}}+b \sqrt{A Z_{c} Z_{d}}+c.
		\end{equation}
UDL can also be seen as a corrected version of it, with multiplying an extra factor $\sqrt{A_{c}^{1 / 3}+A_{d}^{1 / 3}}$ to the second term. Predictive residuals of the three laws are plotted in Fig. \ref{3lawscomparison_ee}. It shows that the uncorrected formula tends to overestimate the nuclei with $\text{N}> 126$ and underestimate the nuclei with $\text{N}\leq126$. Both the corrections laws, NGNL and UDL, cause the shifts towards smaller residues in Fig. \ref{3lawscomparison_ee}. Comparing between two laws, the term S in NGNL yields a more homogeneous distribution when crossing the $N=126$ shell closure. UDL gives correct but little modifications to the uncorrected formula.
	
	Interestingly, UDL much improves the description on nuclei just above the $N=126$ shell closure compared with the uncorrected formula. One possible explanation can be found from the explicit form of parameter $b$. Based on different but equivalent physical consideration, UDL gives $\displaystyle b=-\frac{2 e \sqrt{2 m_{n} r_{0}}}{\hbar \ln 10}+ const $ and the uncorrected formula gives $\displaystyle b=-\frac{4 e \sqrt{2 m_{n} R}}{\hbar \ln 10} + const$ \cite{ni2008unified}. The difference appears merely on the factor corresponds to the nuclear radius, while UDL assumes $R=r_{0}\left(A_{c}^{1 / 3}+A_{d}^{1 / 3}\right)$ \cite{qi2009microscopic} and the uncorrected formula regards it as an global average property \cite{ni2008unified}.

	The preformation radius is normally estimated via the Bohr-Sommerfeld quantization condition. A preformed $\alpha$ cluster in quasi-bound state locates on an orbit, of which radius $R$ satisfies:
	\begin{equation}
		\int_{0}^{R} \sqrt{2 \mu\left(E_{\alpha}+V_{N}-\frac{Z_{c} Z_{d} e^{2}}{4\pi \epsilon R}\right)} d r=\frac{\pi \hbar}{2}(G+1),
	\end{equation}
where the global quanta $G=2n+L$ and where $\mu$ denotes the reduced mass. The orbit radius is estimated by the following quadratic formula with a constant nuclear potential $V_{N}$ as a simple estimation:
	\begin{equation}
		\begin{aligned}
		\displaystyle
		R&=\frac{ \frac{Z_{c} Z_{d} e^{2}}{4\pi \epsilon}+\sqrt{\left(\frac{Z_{c} Z_{d} e^{2}}{4\pi \epsilon}\right)^{2}+\frac{\pi^{2} \hbar^{2}}{2 \mu}\left(E_{\alpha}+V_{N}\right)(G+1)^{2}}}{2\left(E_{\alpha}+V_{N}\right)} 	\\
		&\approx \frac{Z_{c} Z_{d} e^{2}}{8\pi \epsilon\left(E_{\alpha}+V_{N}\right)}+\frac{\pi \hbar(G+1)}{\sqrt{8 \mu\left(E_{\alpha}+V_{N}\right)}}.
		\end{aligned}
	\end{equation}
The approximation is reasonable since the comparison:
	\begin{equation}
		\frac{\pi^{2}}{2} \frac{Q+V_{N}}{\mu c^{2}}\left(\frac{G+1}{Z_{c} Z_{d}} \frac{4\pi \epsilon \hbar c}{e^{2}}\right)^{2} \approx 47.8483
	\end{equation}
holds. Buck pointed out that the global quanta increases from 22 to 24 after crossing the shell closure \cite{buck1991ground}. The result is a deviation of the preformation radius:
	\begin{equation}
		\Delta R=\frac{\pi \hbar}{\sqrt{2 m_{n} A\left(E_{\alpha}+V_{N}\right)}},
	\end{equation}
where $\mu = m_{n}A$ and $m_{n}$ denotes the nucleon mass. If a form $\displaystyle b \propto \frac{2 e \sqrt{2 m_{n} R}}{\hbar \ln 10}$ is taken, the contribution of $\Delta R$ on the residue of half-life can be estimated:
	\begin{equation}
		\Delta = \Delta b \sqrt{A Z_{c} Z_{d}} = - \frac{2 e \pi}{\ln 10} \sqrt{\frac{Z_{c} Z_{d}}{R\left(E_{\alpha}+V_{N}\right)}}
	\end{equation}
Within the errors of our treatment we take $m_{n}c^{2} \approx 938.9 \text{MeV}$, $\hbar c = 197.3 \text{MeV}$ and the nuclear potential $V_{N} \approx 100 \text{MeV}$. The numerical result shows that $\Delta \approx -0.68$, which gives quite similar values to the term S.
	
It is of interest to replace the constant S in NGNL with the explicit form of the radius. To reduce the complexity, the radius on the denominator is estimated via $R = r_{0} \left( A_{c}^{1/3} + A_{d}^{1/3} \right)$. Hence, the corrected formula reads:
	\begin{equation}
    \label{corrected}
	\begin{aligned}
		\log T_{1 / 2}&=a Z_{c} Z_{d} \sqrt{\frac{\mu}{Q_{\alpha}}}+b \sqrt{\mu Z_{c} Z_{d}}+ c \\ &+ d \delta_{N > 126} \sqrt{\frac{Z_{c} Z_{d}}{\left(A_{c}^{1 / 3}+A_{d}^{1/3}\right)(E_{\alpha}+100)}},
	\end{aligned}
	\end{equation}
where $\delta_{N > 126}$ indicates that the correction only validates for nuclides beyond $N=126$ shell closure. The new corrected formula depends on 4 parameters. As shown in TABLE \ref{param_digit_ee}, NGNL and the corrected formula give a quite similar global prediction, which shows similar results from term S in NGNL and the present corrected term. A better description on radius may improve the results and could be further investigated in the future.

\begin{figure*}[]
	\includegraphics[width=0.95\textwidth]{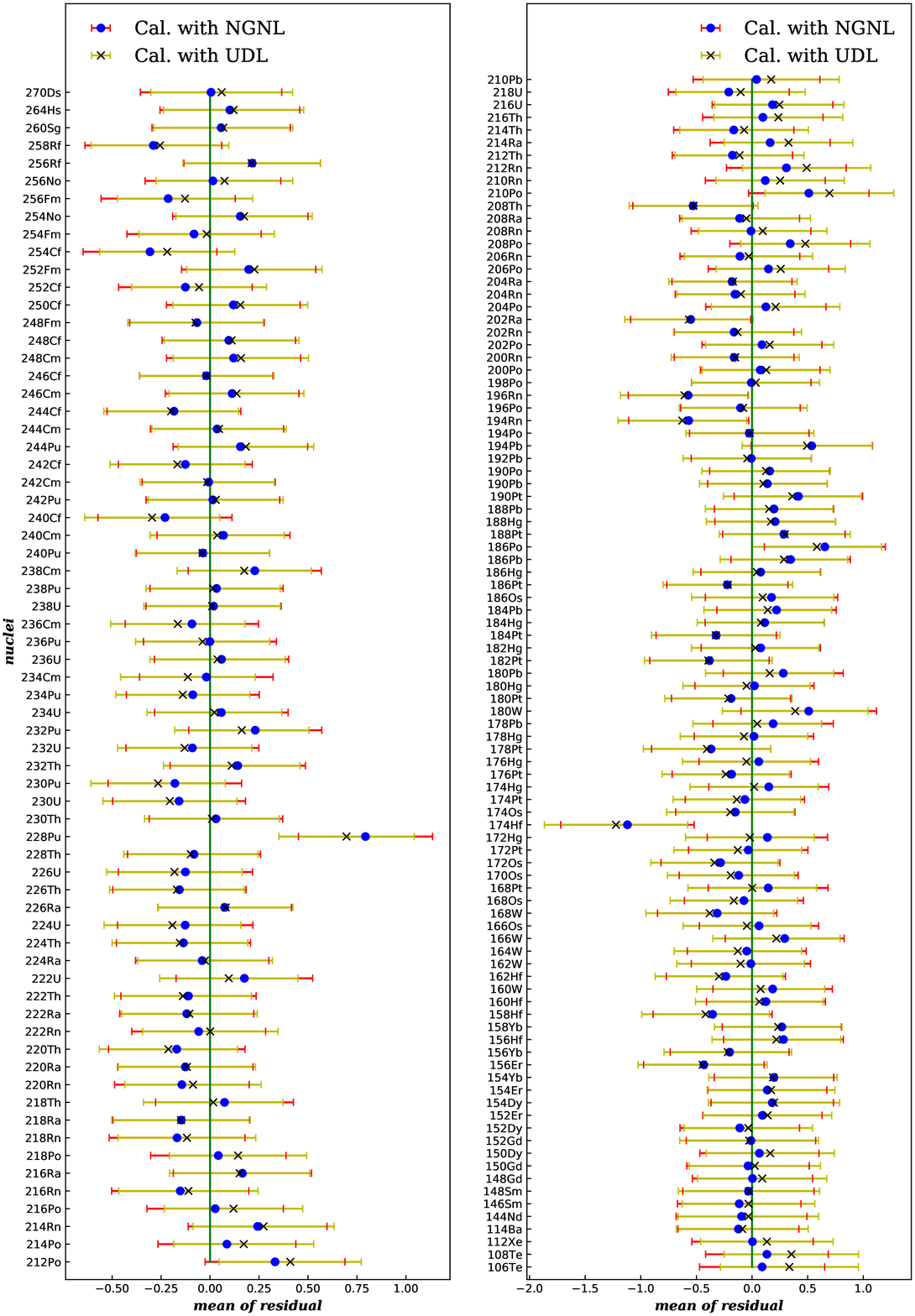}
	\caption{\label{fig:errorbar}Error bar for each even-even nucleus with $N > 126$ (left) and $N \leq 126 $ (right) by UDL and NGNL. Point value is the mean of residual and bar width takes 2$\sqrt{\sigma^{2}_{sys}+\sigma^{2}_{stat}(Nu_{i})}$.}
\end{figure*}

\begin{table*}[]
\caption{Parameters and uncertainties obtained from Gs-Ex channels and Gs-Gs channels of corresponding nuclei for both UDL and NGNL and changes of formation probability.}
\label{param_digit_gsgs_gsex}
	\begin{threeparttable}
		\begin{ruledtabular}
			\begin{tabular}{cccccccccccccc}
			 &  & \multicolumn{2}{c}{a} & \multicolumn{2}{c}{b} & \multicolumn{2}{c}{c} & \multicolumn{2}{c}{P} & \multicolumn{3}{c}{Uncertainty} & $\overline{\text{FP}_{Gs-Ex}/\text{FP}_{Gs-Gs}}$ \\
			 &  & Mean & S.D. & Mean & S.D. & Mean & S.D. & Mean & S.D. & total & stat & sys & \\
			\colrule
			\multirow{2}{*}{UDL} & Gs-Gs & 0.4071 & 0.0011 & -0.3556 & 0.0087 & -26.7438 & 0.6287 & \textemdash   & \textemdash  & 0.1077 & 0.0287 & 0.1038 & \textemdash   \\
			                                 & Gs-Ex & 0.4054 & 0.0020 & -0.2677 & 0.0138 & -33.0907 & 1.2068 & 0.0553\tnote{*} & 0.0056\tnote{*} & 0.1407 & 0.0439 & 0.1336 & 0.6197 \\
			\colrule
			\multirow{2}{*}{NGNL} & Gs-Gs & 0.4034 & 0.0009 & -1.2413 & 0.0285 & -19.4226 & 0.7371 & \textemdash   & \textemdash  & 0.1054 & 0.0268 & 0.1019  & \textemdash  \\
			                                    & Gs-Ex & 0.4025 & 0.0019 & -0.9424 & 0.0448 & -27.3417 & 1.3753 & 0.0558\tnote{*} & 0.0053\tnote{*} & 0.1286 & 0.0403 & 0.1222  & 0.7928
			\end{tabular}%
		\end{ruledtabular}
		\begin{tablenotes}
			\footnotesize
			\item[*] Obtained from the union of Gs-Ex channels and corresponding Gs-Gs channels
		\end{tablenotes}
	\end{threeparttable}
\end{table*}

	The analysis we had so far is limited to the transition from the ground state to ground state, which is the dominant channel for the $\alpha$ decays of ee nuclei. The decays from the ground state of mother nuclei to the excited states of daughter nuclei can also be significant but more complicated. Such decays often involve non s-wave emission. The $\alpha$ preformation property is different from that of the dominant emission. As mentioned above, we have included in our database $42$ Gs-Ex channels of ee nuclei with $N > 126$ and $l = 2$. The Gs-Gs channels in the same nuclei are also considered. The comparison between the fitting results of the two channels should provide information on the difference between the $\alpha$ formation probability of the channels. The formation probability (FP) measures the overlap between the $\alpha$ decay state in the mother nucleus and the corresponding final state in the daughter nucleus \cite{qi2019recent}. The value of FP for a given decay channel can be extracted from experimental decay half-lives by subtracting the contribution from the Coulomb and centrifugal barriers.

In both two empirical laws, FP is assumed to be of a linear form (see discussions in Refs. \cite{qi2009universal,qi2009microscopic,ren2012new})
	\begin{equation}
		\log \text{FP} = b_{1} \rho + c_{1},
	\end{equation}
with $b_{1}$, $c_{1}$ two constants, and $\rho$ the corresponding variable to parameter $b$ for certain nucleus in Eq. (\ref{udl}) and (\ref{ngnl}). As mentioned above, in both two laws, $b_{1}$ and $c_{1}$ terms are absorbed in the $b$ and $c$ terms, respectively. It should be noted that an increment on FP will induce the decrement of half-life. The $b_{1}$ term should have an opposite sign to the $b$ term. Assuming the separation radius changes little from decaying to ground state to decaying to excited state, The change of FP can be estimated via
	\begin{equation}\label{FP}
		\log(\text{FP}_{Gs-Ex}/\text{FP}_{Gs-Gs}) =  (\Delta b_{1}) \rho + (\Delta c_{1}).
	\end{equation}
where $\Delta b_{1}=-\Delta b=-(b_{Gs-Ex}-b_{Gs-Gs})$, $\Delta c_{1}=-\Delta c=-(c_{Gs-Ex}-c_{Gs-Gs})$. The change on FP, $\text{FP}_{Gs-Ex}/\text{FP}_{Gs-Gs}$, in each nuclei can be estimated by Eq. (\ref{FP}). Then the average $\overline{\text{FP}_{Gs-Ex}/\text{FP}_{Gs-Gs}}$ for considered 42 nuclei is shown in the TABLE \ref{param_digit_gsgs_gsex}, which is smaller than one and indicates the hindered decay process for Gs-Ex channels.

\subsection{\label{OD}Discussions on Outliers}

\begin{figure*}[]
	\includegraphics[width=.3\textwidth]{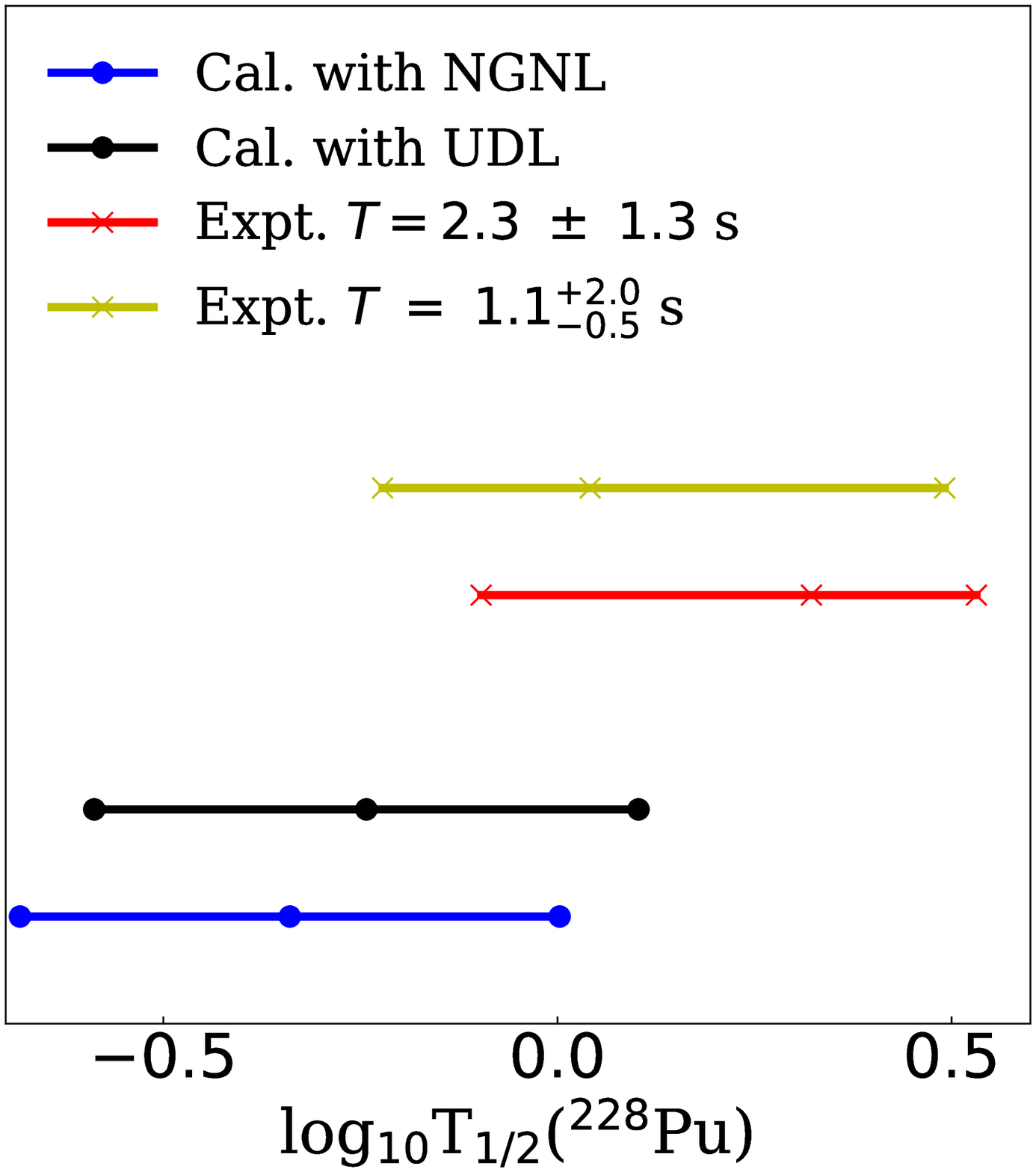}
	\includegraphics[width=.3\textwidth]{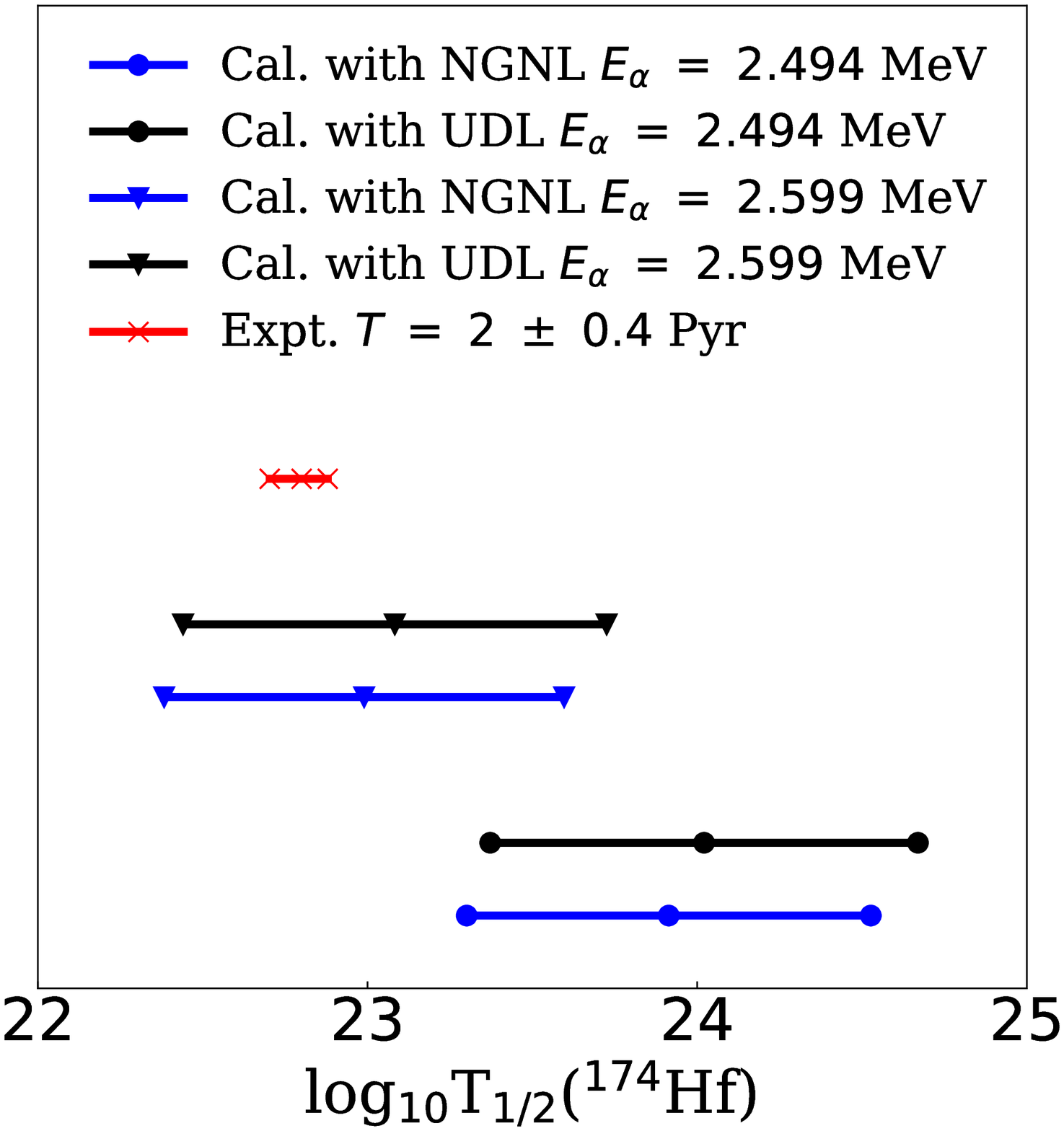}
	\includegraphics[width=.3\textwidth]{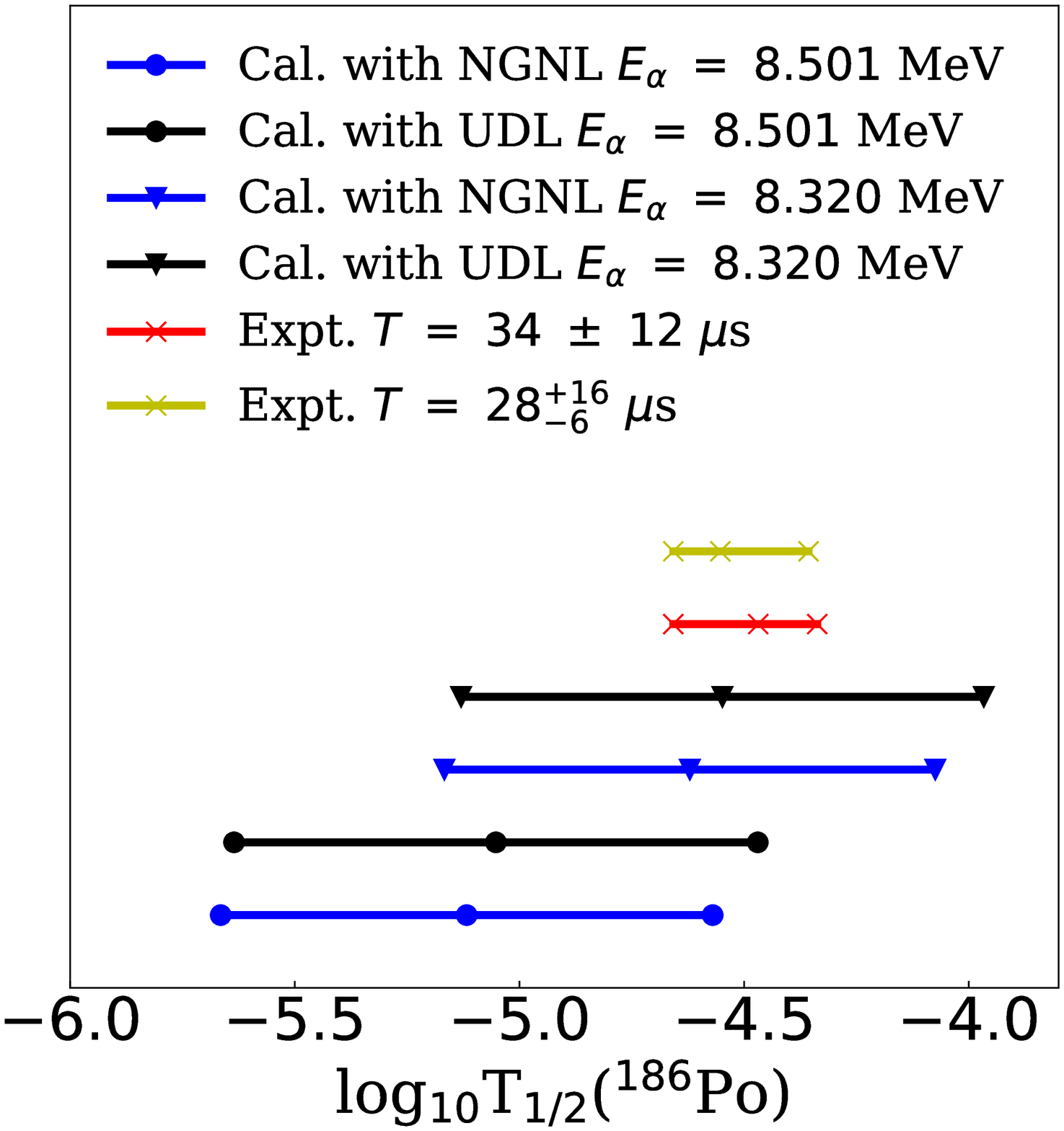}
	\caption{Error bar for $^{228}$Pu (left), $^{174}$Hf (middle) and $^{186}$Po (right). The bar width is 2$\sigma_{total}(Nu_{k})$. The half-lives are given in seconds.}
 	\label{fig:errorbarPuHfPo}
\end{figure*}

	For a given decay formula, some half-lives may be calculated far from the observed data. Theoretical results are presented with error bars in Fig. \ref{fig:errorbar} to identify possible outliers. Note that the parameters taken are deduced from the corresponding nuclei dataset. $^{228}$Pu and $^{174}$Hf are outliers in both two laws, which are identified by the large differences (much larger than 2$\sqrt{\sigma^{2}_{sys}+\sigma^{2}_{stat}(Nu_{i})}$) between their calculated and observed half-lives as seen in Fig. \ref{fig:errorbar}.

For $^{228}$Pu, the observed half-life has a large uncertainty, which is $2.3 \pm 1.3$ s in NUBASE2016 \cite{audi2017nubase2016} and $1.1^{+2.0}_{-0.5}$ s in Ref.~\cite{nishio2003half}, respectively. If observed uncertainties are taken into account, calculated results and observed data agree with each other, as seen in Fig. \ref{fig:errorbarPuHfPo}.

The $2494.52 \pm 2.26 $ keV $\alpha$ decay energy of $^{174}$Hf in AME2016 \cite{wang2017ame2016} is consistent with the one recommended in Ref.~\cite{akovali1998review}, but the same reference recorded another value $2559 \pm 30 $ keV, which is coherent with the radii systematic. If the latter decay energy is used, the calculated results agree rather well with the observed data, as seen in Fig. \ref{fig:errorbarPuHfPo}.

As seen in Fig. \ref{fig:errorbar}, $^{186}$Po is slightly underestimated by both laws. NUBASE2016 \cite{audi2017nubase2016} and AME2016 \cite{wang2017ame2016} gave its half-life and decay energy $34 \pm 12~\mu$s and $8501.17 \pm 13.71$ keV, respectively, while $28^{+16}_{-6}~\mu$s and 8320 $\pm$ 15 keV are respectively presented in Ref.~\cite{andreyev2013signatures}. Comparisons in Fig. \ref{fig:errorbarPuHfPo} indicate calculated results with $8320 \pm 15 $keV are more closed to the observed decay half-lives.

\begin{table}[]
\caption{\label{tab:Te} Experimental data of $^{108}$Xe and $^{104}$Te. }
\begin{ruledtabular}
\begin{tabular}{ccc}
Nuclide & $Q_{\alpha}$ (MeV) & $T_{1/2}$            \\
\colrule
$^{108}$Xe\cite{auranen2018superallowed}   & 4.4(0.2) & $58^{+106}_{-23}$  $\mu$s    \\
$^{104}$Te\cite{auranen2018superallowed}   & 4.9(0.2) & \textless 18 ns	\\
$^{104}$Te\cite{xiao2019search}   &\textemdash & \textless 4 ns	\\
\end{tabular}
\end{ruledtabular}
\end{table}

\begin{figure*}[]
	\includegraphics[width=0.3\textwidth]{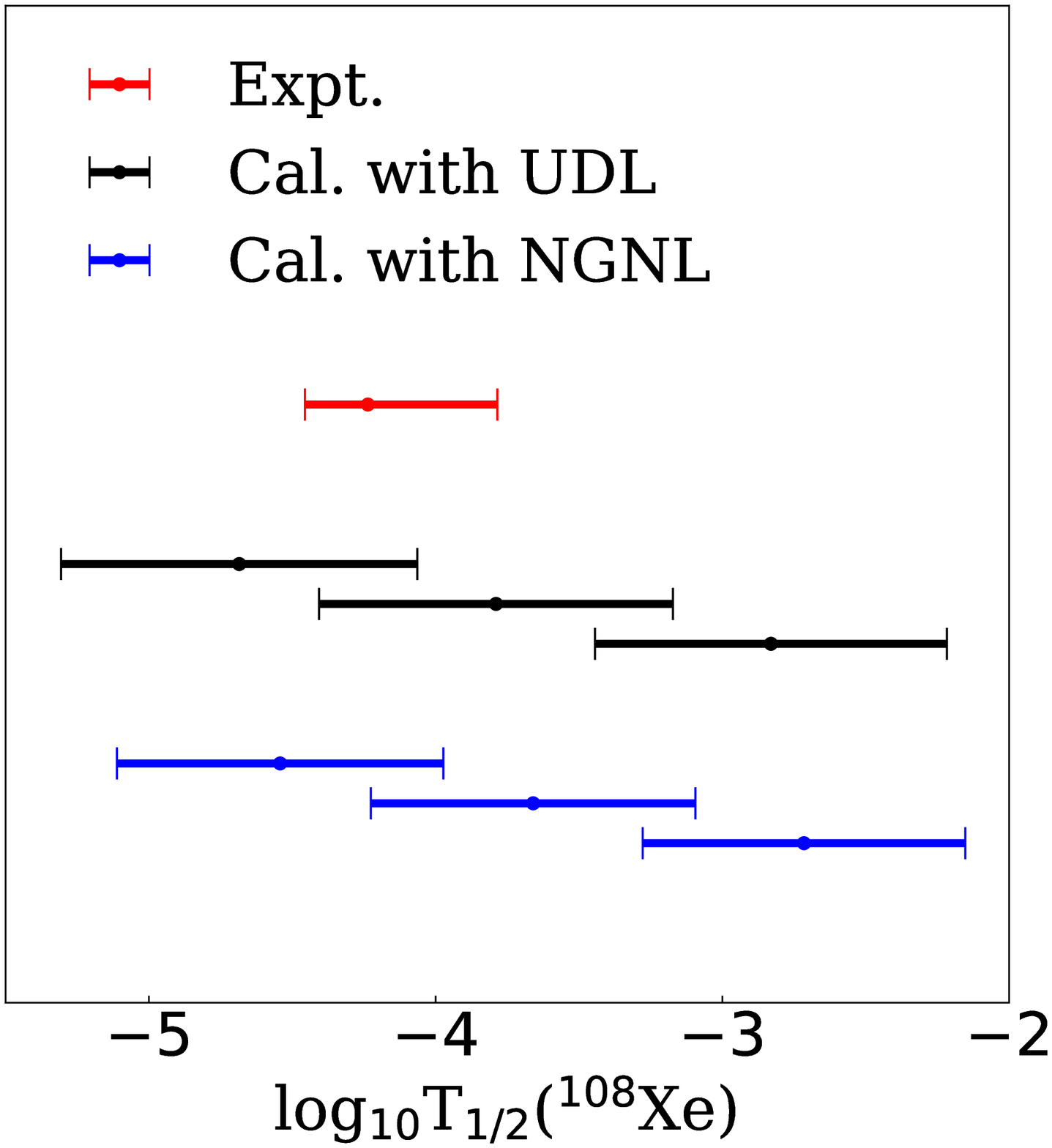}
	\includegraphics[width=0.3\textwidth]{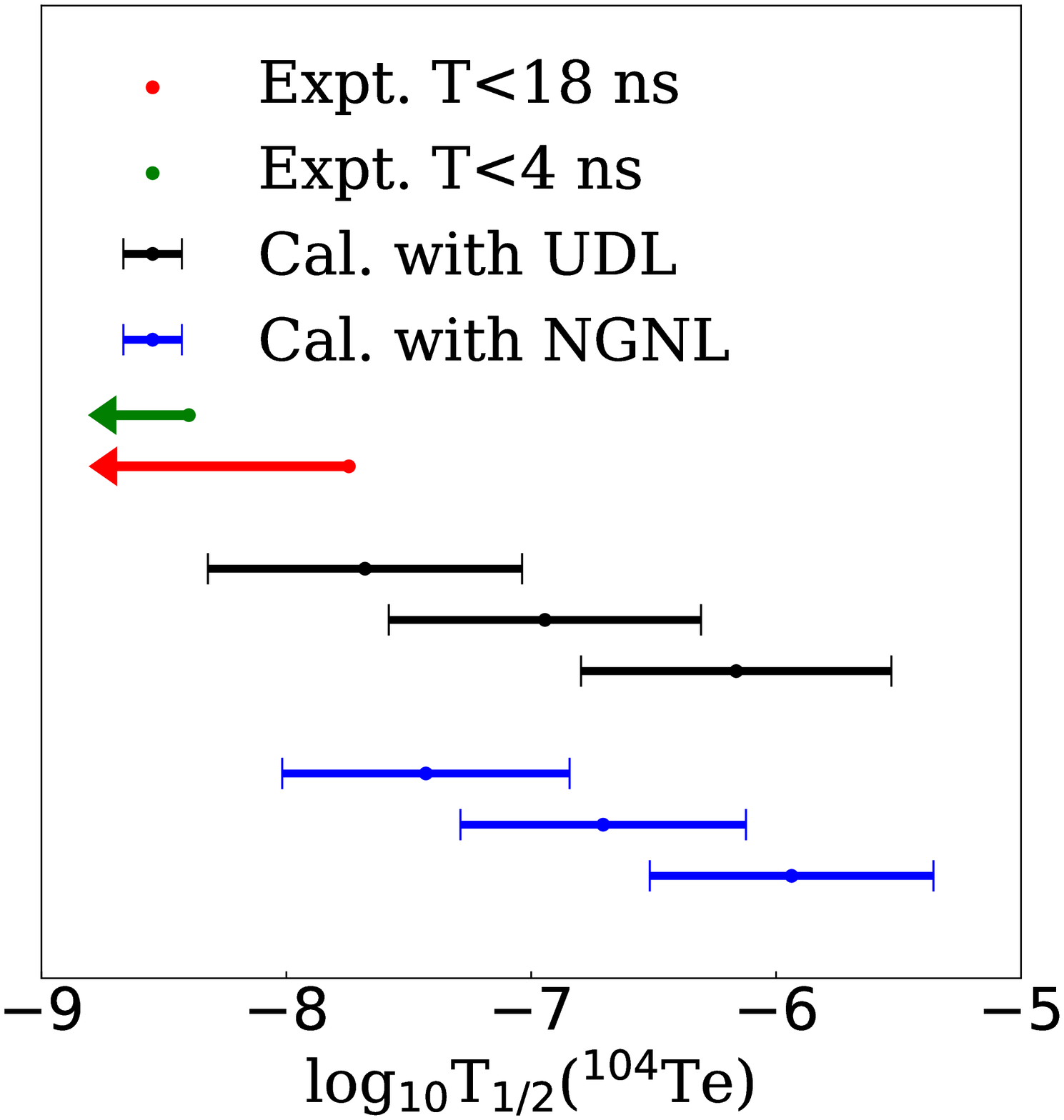}
	\caption{\label{fig:superallowed}Half-lives calculated by UDL and NGNL for decays of $^{108}$Xe $\rightarrow$ $^{104}$Te (left panel) and $^{104}$Te $\rightarrow$ $^{100}$Sn (right panel) using experimental decay energy \cite{auranen2018superallowed}. The green line invokes the experimental results. For $^{104}$Te $\rightarrow$ $^{100}$Sn decay, only the upper bound of half-life is available. The bar width is 2$\sigma_{total}(Nu_{k})$. Predictions are presented with decay energy (middle point), its upper bound (left point) and lower bound (right point). The half-lives are given in seconds.}
\end{figure*}

Another interesting application is on the newly observed superallowed $\alpha$ decay from $^{104}$Te to $^{100}$Sn. A decay chain is newly observed from $^{108}$Xe $\rightarrow$ $^{104}$Te $\rightarrow$ $^{100}$Sn \cite{auranen2018superallowed}. Since $^{100}$Sn is a doubly magic nucleus with the same numbers of protons and neutrons, it is expected that the decay toward such a nucleus possesses an extremely large FP. If the superallowed nature is assumed for the $\alpha$ decay of $^{104}$Te, models with parameters obtained from other nuclei should much overestimate the half-life of $^{104}$Te.

The newly observed half-lives of two decays are compared with theoretical calculations in Fig. \ref{fig:superallowed}. The parameters used for the results in Fig. \ref{fig:superallowed} are obtained from ee nuclei with $N\leqslant126$. The experimental decay energies and half-lives are listed in Table \ref{tab:Te}. The error bars in Fig. \ref{fig:superallowed} show the theoretical uncertainties obtained through the present work, while the center values are calculated with the decay energies and their upper and lower bounds given in Ref.~\cite{auranen2018superallowed}. The left figure suggests that decay of $^{108}$Xe shows no deviations from the global laws with parameters obtained from the other nuclei. A similar conclusion was drawn in Ref. \cite{qi2019recent}. The half-life of $^{104}$Te $\rightarrow$ $^{100}$Sn decay locates outside the theoretical error bars of both laws with experimental decay energy. However, if the upper bound of experimental decay energy is used for calculation, the calculated half-life locates inside the error bars of both laws.

A very recent experiment suggested such an upper bound could be reduced to 4 ns \cite{xiao2019search}. The green arrow in Fig. \ref{fig:superallowed} indicates that the newly measured limit makes superallowed nature more significant compared with the theoretical predictions. However, the superallowed nature may not be properly approved with high confidence (up to 95\%) at this stage by considering the large theoretical and experimental uncertainties.  In addition, one cannot exclude the possibility that additional term may be needed in our present formula for a more proper description of those light nuclei, which will further increase the theoretical uncertainty.

%
%
%
%
\section{\label{AON}Application to odd-mass nuclei}

\begin{figure*}[]
	\includegraphics[width=.24\textwidth,height=.24\textwidth]{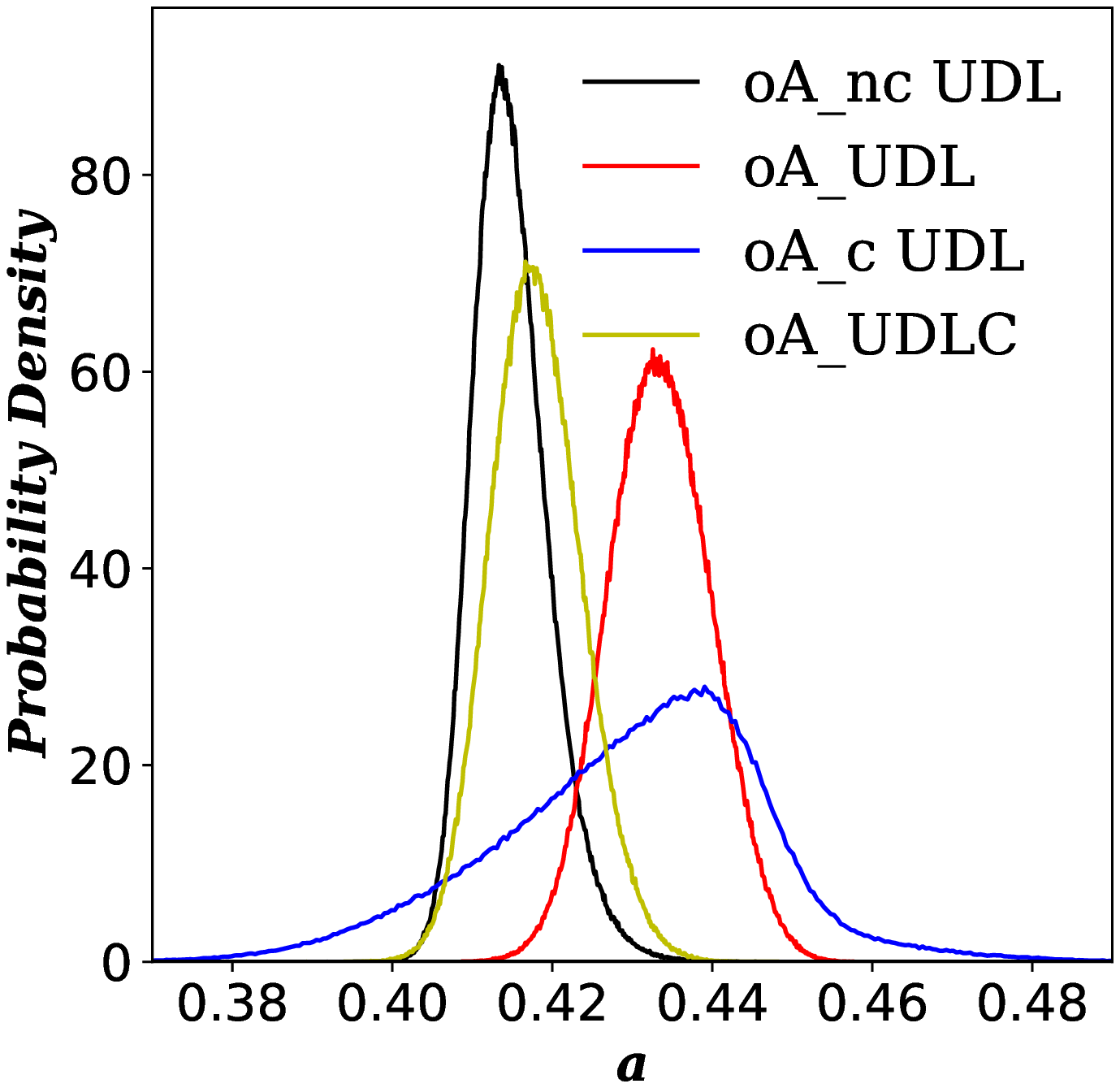}
	\includegraphics[width=.24\textwidth,height=.24\textwidth]{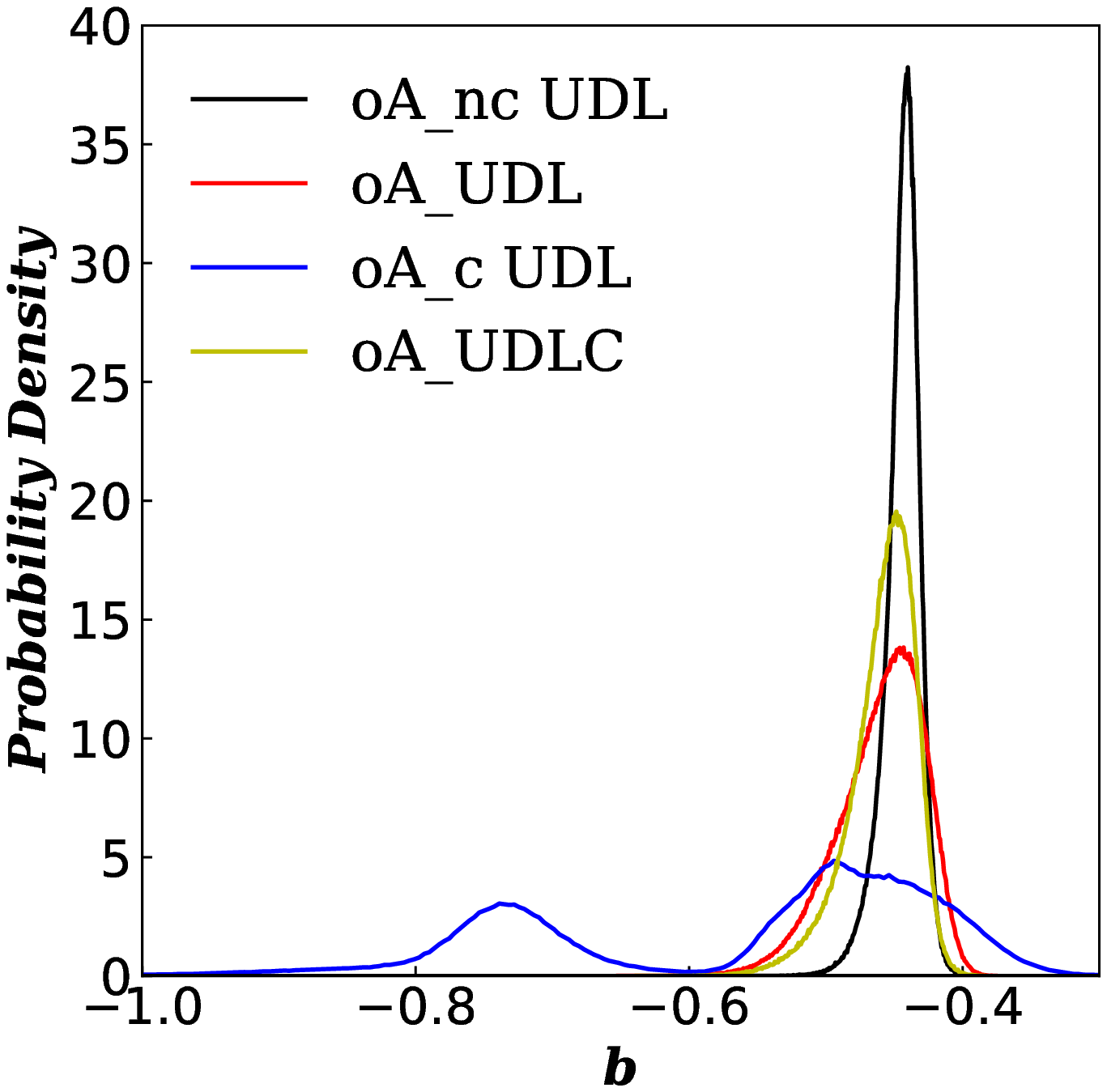}
	\includegraphics[width=.24\textwidth,height=.24\textwidth]{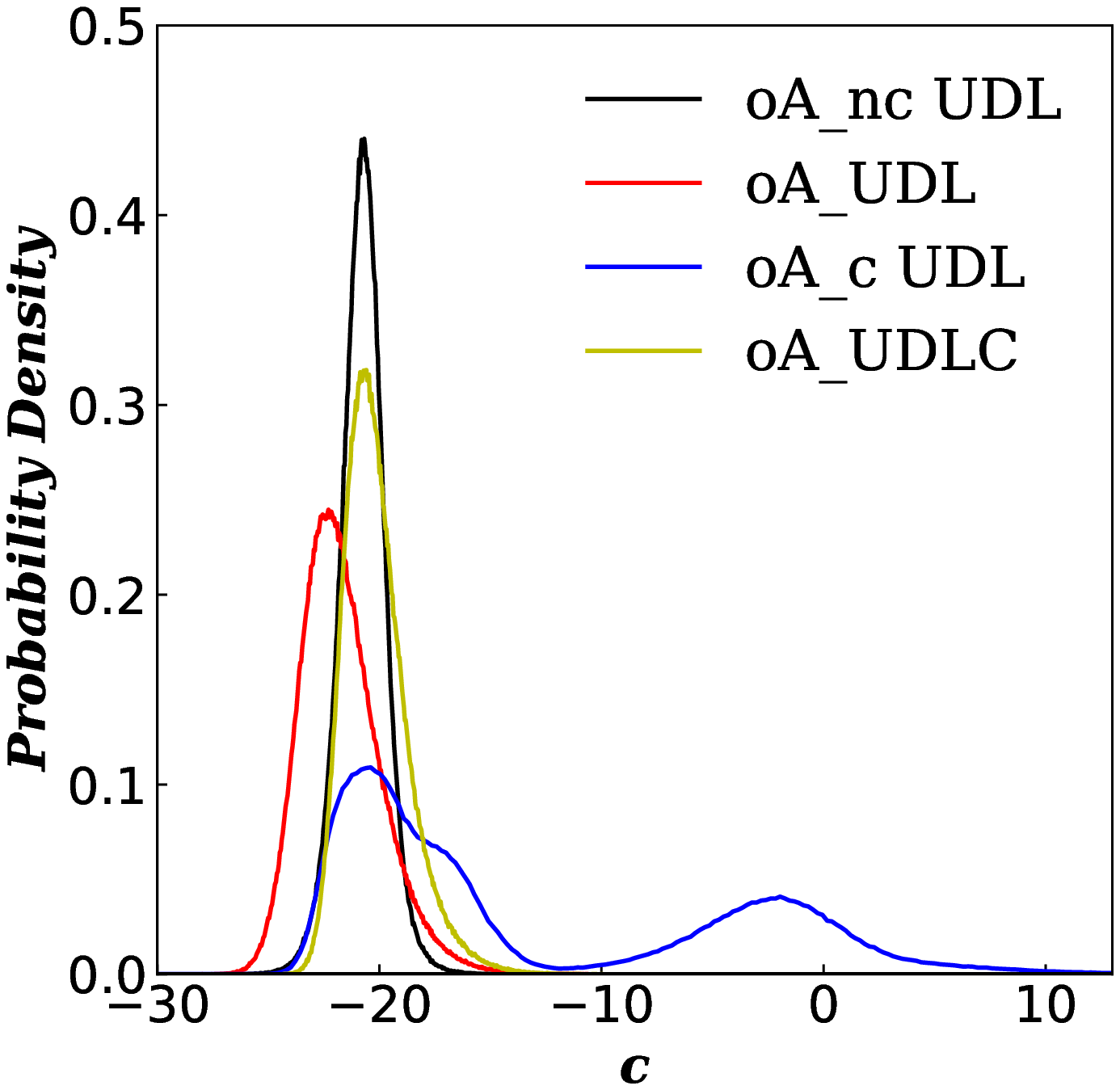}
	\includegraphics[width=.24\textwidth,height=.24\textwidth]{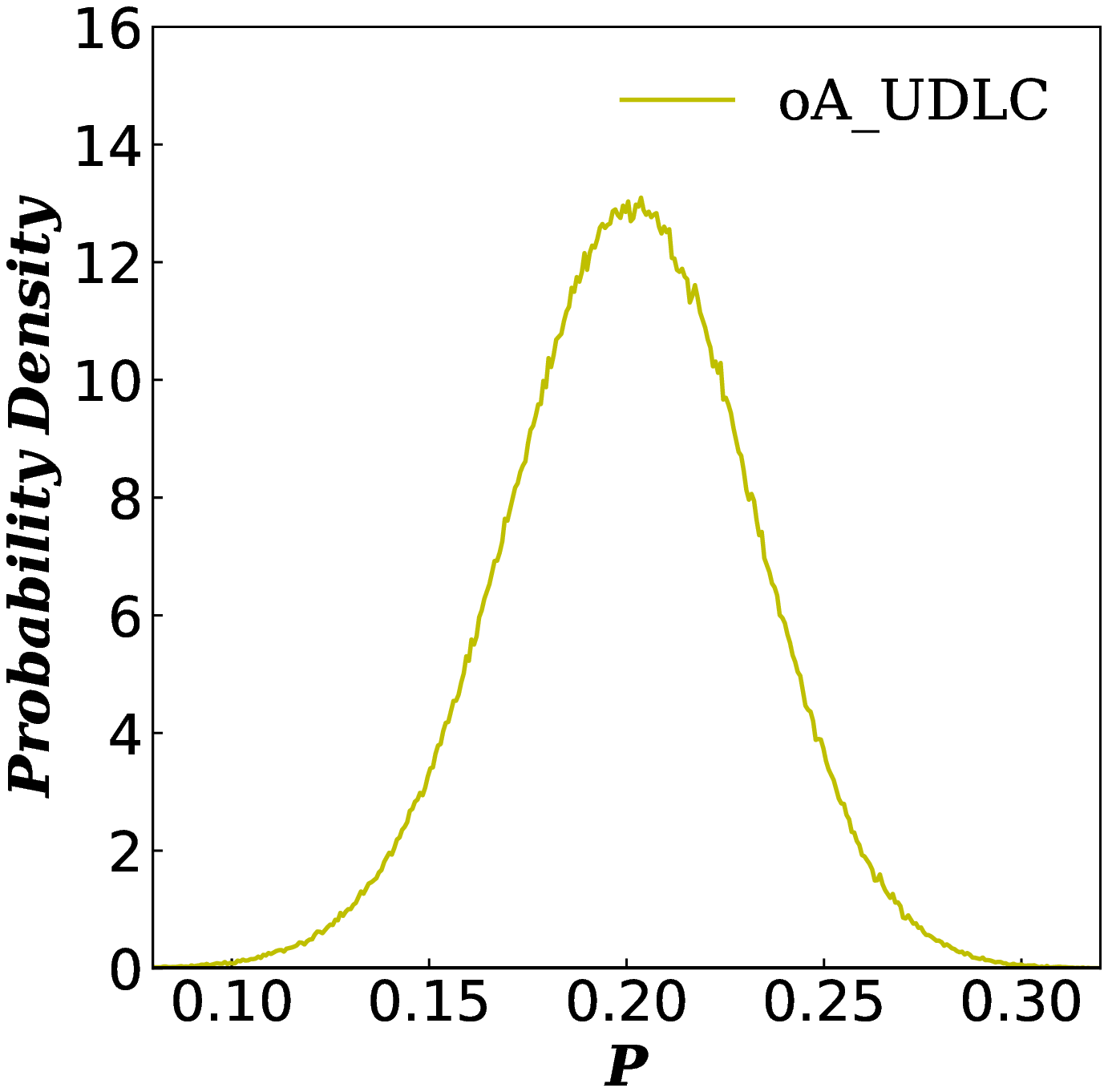} \\
	\includegraphics[width=.24\textwidth,height=.24\textwidth]{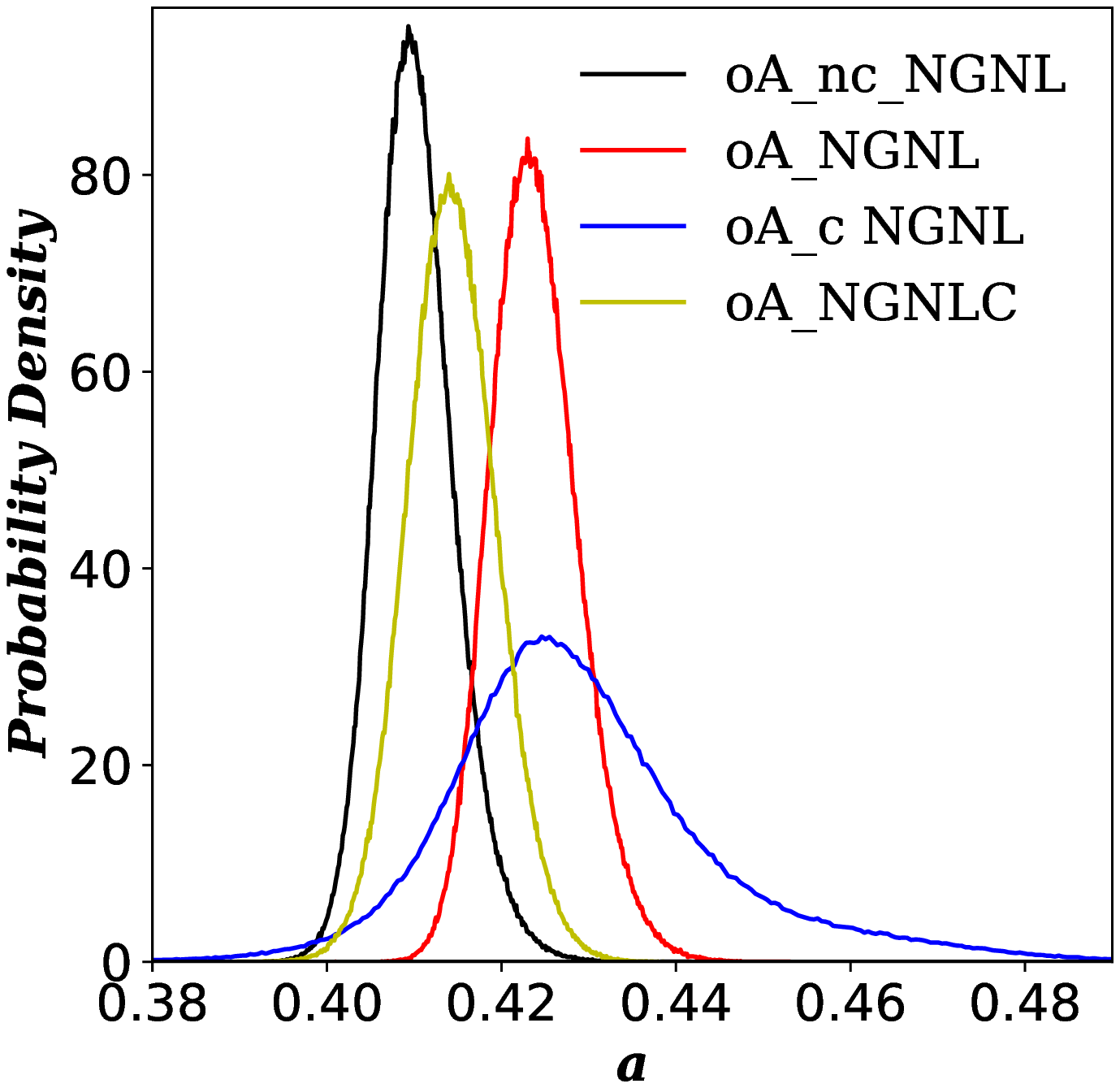}
	\includegraphics[width=.24\textwidth,height=.24\textwidth]{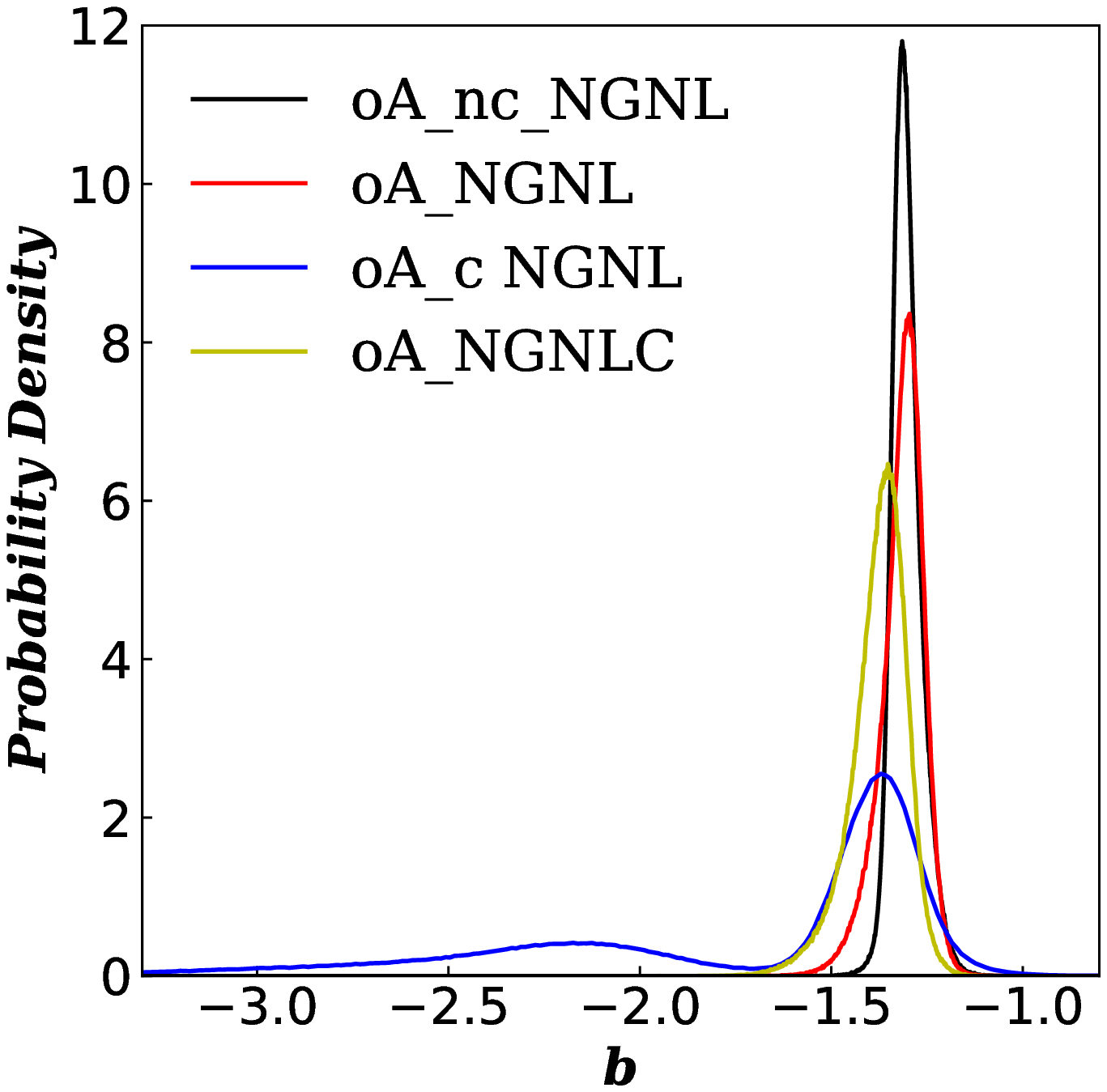}
	\includegraphics[width=.24\textwidth,height=.24\textwidth]{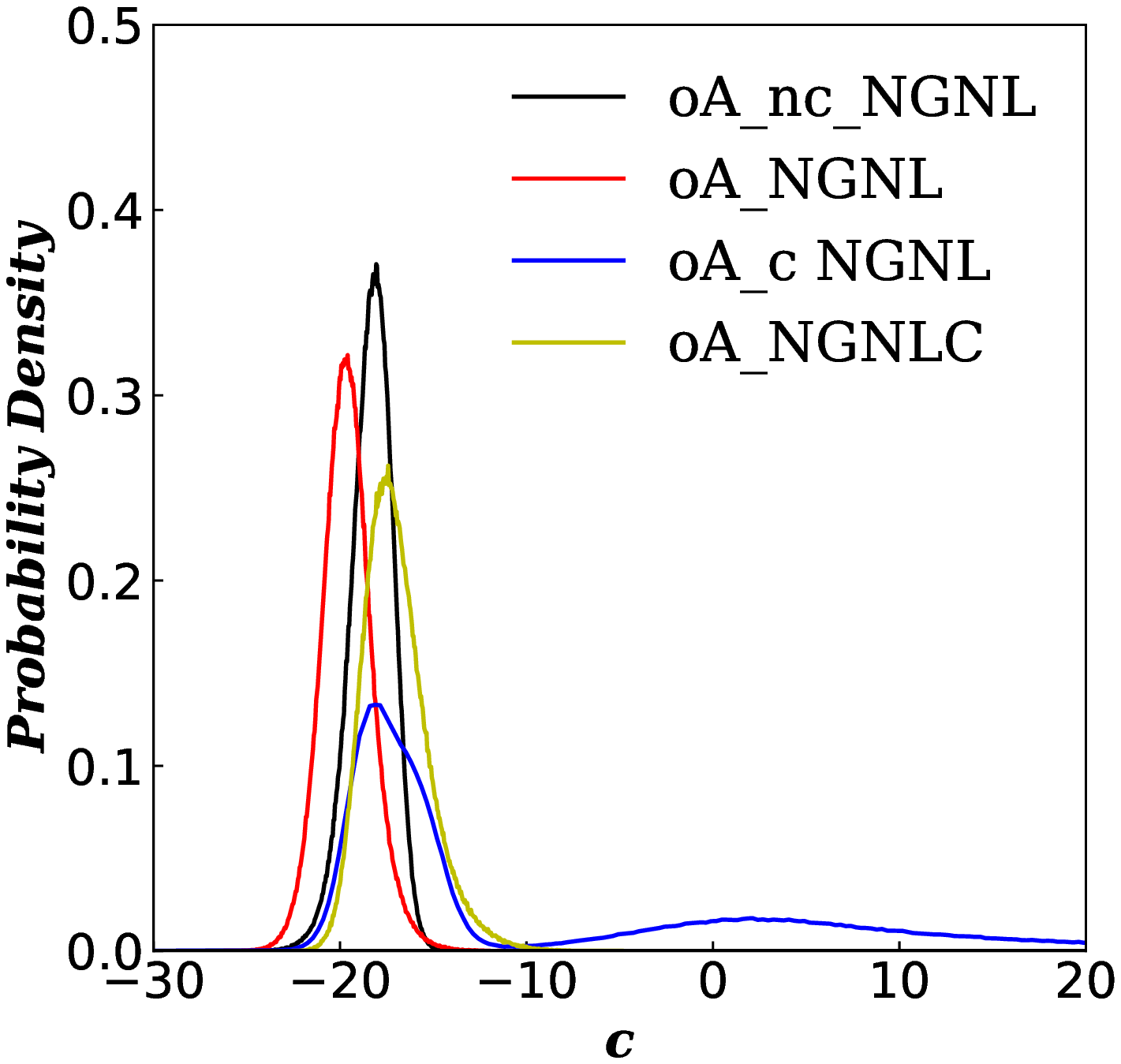}
	\includegraphics[width=.24\textwidth,height=.24\textwidth]{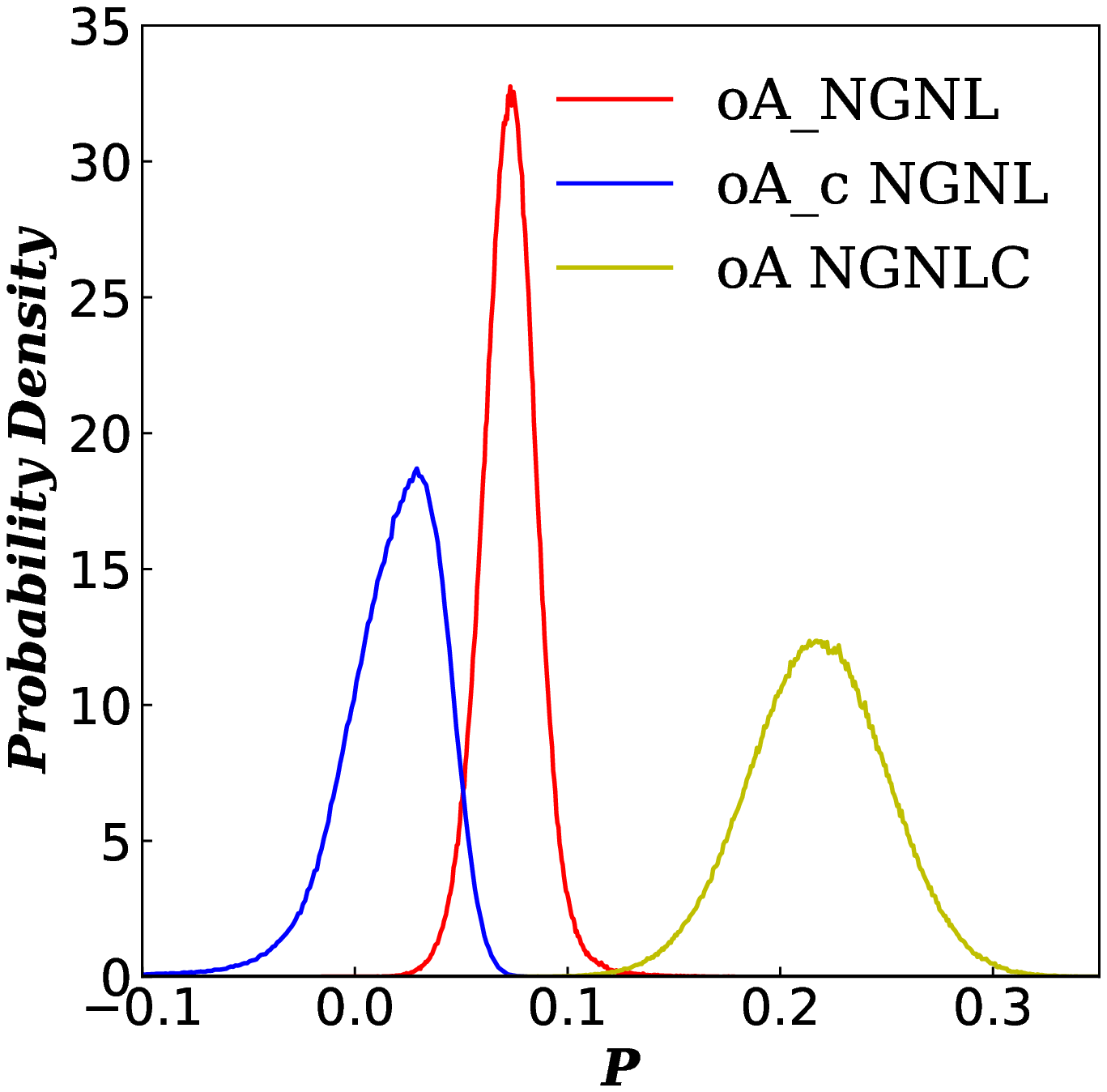} \\
	\caption{\label{fig:dis} Distributions of parameters for UDL (top) , UDLC (top), NGNL (bottom), and NGNLC (bottom) fitted to odd mass nuclei. From left to right, three figures invoke parameters $a$, $b$, $c$, and $P$ in turn.}
\end{figure*}

\begin{table*}[]
\caption{\label{tab:stat}Parameters and decomposed uncertainties obtained from ee nuclei, oA nuclei, oA\_c nuclei, and oA\_nc nuclei for UDL, UDLC, NGNL, and NGNLC.}
\begin{threeparttable}
\begin{ruledtabular}
\begin{tabular}{ccccccccccccccc}
            &       & \multicolumn{2}{c}{$a$} & \multicolumn{2}{c}{$b$} & \multicolumn{2}{c}{$c$}      & \multicolumn{2}{c}{$P$}          &  Residuals&       \multicolumn{3}{c}{Uncertainties}          \\
            \colrule
            &       & Mean       & S.D.       & Mean  & S.D.     & Mean    & S.D.    & Mean    & S.D.                                                        &mean        & $\sigma_{total}$ & $\sigma_{stat}$ & $\sigma_{sys}$  \\
            \colrule
\multirow{6}{*}{UDL}     & oA     & 0.4334 & 0.0064    & -0.4600 & 0.0317     & -21.7149 & 1.8186       &  \textemdash  &  \textemdash  & \textemdash & 0.8878   & 0.1648   & 0.8723          \\
            & oA\_c  & 0.4299  & 0.0168     & -0.5620  & 0.1465        & -13.2482 & 8.8444       &  \textemdash  &  \textemdash                           & \textemdash & 0.9477   & 0.4083  & 0.8552           \\
            & oA\_nc & 0.4149 & 0.0048  & -0.4442   & 0.0134   & -20.7585 & 1.0332      &  \textemdash  &  \textemdash                           & \textemdash & 0.5427   & 0.1283  & 0.5273            \\
            & oA\_nc\tnote{T} & 0.4150 & 0.0048  & -0.4445   & 0.0137   & -20.7557 & 1.0421      &  \textemdash  &  \textemdash                           & \textemdash & 0.5429   & 0.1291  & 0.5273            \\
            & oA\_nc\tnote{E} & 0.4149 & 0.0048  & -0.4441   & 0.0141   & -20.7614 & 1.0704      &  \textemdash  &  \textemdash                           & \textemdash & 0.5431   & 0.1298  & 0.5273            \\
            & ee     & 0.4096 & 0.0022    & -0.4266 & 0.0043   & -21.5675 & 0.2880   &  \textemdash  &  \textemdash                          & \textemdash & 0.3442   & 0.0503  & 0.3405            \\
            &oA\_nc\tnote{1}   & \textemdash & \textemdash & \textemdash & \textemdash & \textemdash & \textemdash & \textemdash & \textemdash                                                                                                                                                  &0.3544     & 0.6369 & 0.0425 & 0.6355\\
            &oA\_c\tnote{2}    & \textemdash & \textemdash & \textemdash & \textemdash & \textemdash & \textemdash & \textemdash & \textemdash                                                                                                                                                   &1.2534     & 1.5067 & 0.2209 & 1.4905 \\
UDLC & oA & 0.4180 & 0.0057  & -0.4588 & 0.0249  & -20.1563 & 1.4647  & 0.2004 & 0.0310                                                      & \textemdash & 0.7129   & 0.1502  & 0.6969\\
\colrule
\multirow{8}{*}{NGNL} & oA     & 0.4238 & 0.0050  & -1.3152 & 0.0587    & -19.6084 & 1.3741        & 0.0729           & 0.0136                           & \textemdash & 0.7855   & 0.1516  & 0.7707             \\
               & oA\_c  & 0.4292 & 0.0157   & -1.721 & 0.5477          & -9.0366 & 12.7269        & 0.0173            & 0.0275                                              & \textemdash & 0.9571   & 0.4491  & 0.8452             \\
               & oA\_nc & 0.4103 & 0.0045  & -1.3028 & 0.0397   & -18.3330 & 1.1501      &  \textemdash  &  \textemdash                          & \textemdash & 0.5193   & 0.1204  & 0.5051             \\
               & oA\_nc\tnote{T} & 0.4104 & 0.0046  & -1.3038 & 0.0403   & -18.3252 & 1.1545      &  \textemdash  &  \textemdash                          & \textemdash & 0.5193   & 0.1207  & 0.5051             \\
               & oA\_nc\tnote{E} & 0.4102 & 0.0046  & -1.3024 & 0.0429   & -18.3401 & 1.2116      &  \textemdash  &  \textemdash                          & \textemdash & 0.5197   & 0.1221  & 0.5051             \\
               & ee     & 0.4075 & 0.0019  & -1.3255 & 0.0125     & -17.7229 & 0.2150   &  \textemdash  &  \textemdash                          & \textemdash & 0.2658   & 0.0407  & 0.2626             \\
               &oA\_nc\tnote{1}   & \textemdash & \textemdash & \textemdash & \textemdash & \textemdash & \textemdash & \textemdash & \textemdash                                                                                                                                          &0.3430    & 0.6292 & 0.0391 & 0.6280\\
               &oA\_c\tnote{2}     & \textemdash & \textemdash & \textemdash & \textemdash & \textemdash & \textemdash &0.3092\tnote{3}&0.4056\tnote{3}                                                                                                                                        &1.2573     & 1.5202 & 0.2134 & 1.5051 \\
               &oA\_c($l=2$)\tnote{2}& \textemdash & \textemdash & \textemdash & \textemdash & \textemdash & \textemdash &0.1425\tnote{3}&0.1107\tnote{3}                                                                                                                                        &0.8551    & \textemdash   & \textemdash   &  \textemdash  \\
NGNLC & oA & 0.4144 & 0.0052  & -1.3766 & 0.0748  & -16.9912 & 1.7585  & 0.2175 & 0.0325                                                  & \textemdash & 0.7128  & 0.1478  & 0.6973          \\
\end{tabular}
\end{ruledtabular}
\begin{tablenotes}
\footnotesize
\item[1] oA\_nc estimated by parameters from ee. With consideration of shell effect, $a$ = 0.4013, $b$ = -0.3707, $c$ = -24.7370 for nuclei with $\text{N} > 126$; $a$ = 0.4188, $b$ = -0.3943, $c$ = -24.7439 for nuclei with $\text{N} \leqslant 126$ for UDL. $a$ = 0.3982, $b$ = -1.2622, $c$ = -18.0730 for nuclei with $\text{N} > 126$; $a$ = 0.4126, $b$ = -1.3574, $c$ = -17.6370 for nuclei with $\text{N} \leqslant 126$ for NGNL.
\item[2] oA\_c estimated by parameters from oA\_nc. With consideration of shell effect, $a$ = 0.4189, $b$ = -0.5076, $c$ = -16.8124 for nuclei with $\text{N} > 126$; $a$ = 0.4248, $b$ = -0.3769, $c$ = -26.4414 for nuclei with $\text{N} \leqslant 126$ for UDL. $a$ = 0.4154, $b$ = -1.6855, $c$ = -8.9271 for nuclei with $\text{N} > 126$; $a$ = 0.4112, $b$ = -1.3548, $c$ = -17.3023 for nuclei with $\text{N} \leqslant 126$ for NGNL.
\item[3] $P$ obtained by making the regression to the half-lives residuals of oA\_c nuclei.
\item[T] Bootstrap with consideration of half-life experimental uncertainty.
\item[E] Bootstrap with consideration of $\alpha$ decay energy experimental uncertainty.
\end{tablenotes}
\end{threeparttable}
\end{table*}

\begin{table}
\caption{\label{pearson_oA}Pearson matrix of parameters in UDL, UDLC, NGNL, and NGNLC, fitted with different datasets.}
\begin{ruledtabular}
\begin{tabular}{cc|cccc|cccc}
                              &   & & \multicolumn{2}{c}{UDL} &             &     & \multicolumn{2}{c}{NGNL}& \\
                               &   & $a$       & $b$                  & $c$       & $P$    & $a$              & $b$          & $c$   & $P$ \\
\colrule
\multirow{4}{*}{oA} & $a$ & 1.000 &\textemdash &\textemdash &\textemdash & 1.000 &\textemdash &\textemdash &\textemdash \\
                              & $b$ & -0.643  & 1.000 &\textemdash &\textemdash &-0.395 & 1.000 &\textemdash &\textemdash \\
                              & $c$ & 0.319   & -0.930  & 1.0 &\textemdash & -0.029 &  -0.906  & 1.0  &\textemdash \\
                              & $P$ &\textemdash &\textemdash &\textemdash &\textemdash &  -0.516 &  0.148 & 0.060 & 1.0 \\
\colrule
\multirow{4}{*}{oA\_c}   & $a$ & 1.000 &\textemdash &\textemdash &\textemdash & 1.000 &\textemdash &\textemdash &\textemdash \\
                                     & $b$ & -0.778  & 1.000 &\textemdash &\textemdash & -0.796 & 1.000 &\textemdash &\textemdash \\
                                     & $c$ & 0.651   & -0.983 & 1.000  &\textemdash & 0.723 & -0.993 & 1.000  &\textemdash  \\
                                     & $P$ &\textemdash &\textemdash &\textemdash &\textemdash & -0.446 &  0.525 & -0.527  & 1.0 \\
\colrule
\multirow{4}{*}{oA\_nc}   & $a$ & 1.000 &\textemdash &\textemdash &\textemdash & 1.000 &\textemdash &\textemdash &\textemdash \\
                                       & $b$ & -0.131 & 1.000 &\textemdash &\textemdash & 0.0002 & 1.000 &\textemdash &\textemdash  \\
                                       & $c$ & -0.455   & -0.822 & 1.000  &\textemdash & -0.481 & -0.876 & 1.000  &\textemdash \\
                                       & $P$ &\textemdash &\textemdash &\textemdash &\textemdash &\textemdash &\textemdash &\textemdash &\textemdash \\
\colrule
                                &   & & \multicolumn{2}{c}{UDLC}  &             &    & \multicolumn{2}{c}{NGNLC}& \\
                                &   & $a$       & $b$             & $c$    &$P$    & $a$          & $b$      & $c$   & $P$ \\
\colrule
\multirow{4}{*}{oA}  & $a$ & 1.000 &\textemdash &\textemdash &\textemdash & 1.000 &\textemdash &\textemdash &\textemdash \\
                               & $b$ &-0.557 & 1.000 &\textemdash &\textemdash &-0.401 & 1.000 &\textemdash &\textemdash \\
                               & $c$ &0.162 &-0.909 & 1.000 &\textemdash &0.060  &-0.938 & 1.000 &\textemdash \\
                               & $P$&-0.459 &-0.044 & 0.261 &1.0 & -0.570 &-0.039 &0.244 &1.0 \\

\end{tabular}
\end{ruledtabular}
\end{table}

In UDL, the Coulomb-Hankel function takes the approximation of $H_{0}^{+}(\chi, \rho)$ at the case of $\alpha$ decay between ground states of ee nuclei, of which the spins and parities are both $0^{+}$ \cite{qi2009universal, qi2009microscopic}. As a result, modifications should be considered for UDL to describe the $\alpha$ decay half-life between oA nuclei, as is done in Ref.~\cite{qi2012effects} for proton decay. NGNL proposed a correction for decays with changes of spins \cite{ni2008unified}. In fact, $\alpha$ decays between oA nuclei are different from those between ee nuclei mainly because of the pairing effect and change of spin and/or parity values. The dataset of 92 $\alpha$ decay channels is hence separated into two groups: (i) without spin or parity change (oA\_nc) and (ii) with spin and/or parity change (oA\_c).
	
Parameter distributions obtained with the Bootstrap method are drawn in Fig. \ref{fig:dis}. Table \ref{pearson_oA} presents the correlation coefficients between each pair of parameters, which show strong correlations between parameters $b$ and $c$. The statistics of parameters and the decomposed uncertainties are listed in Table \ref{tab:stat}. Similar to the case of ee nuclei, the results show that the fitting parameters and their uncertainties are barely influenced by the experimental uncertainty. For simplicity, the test is limited to oA\_nc nuclei. From these results, some special discussions can be addressed.

The first one is the parameters obtained from ee nuclei cannot be used to estimate the half-lives of oA\_nc nuclei, although both the ee nuclei and oA\_nc nuclei correspond to $l = 0$. The peak value of the parameter $a$ of ee nuclei presents a negative shift compared with those of the oA\_nc, oA\_c, and oA nuclei. The half-lives of oA\_nc nuclei are globally underestimated by the parameters from ee nuclei, as shown in Table \ref{tab:stat}. Since the first term of both UDL and NGNL corresponds to the quantum tunneling effect, a hindrance exists for oA nuclei during the emission process of an $\alpha$ particle, which is not surprising.

To investigate the difference caused by the unpaired nucleon, parameters obtained with ee nuclei were applied to calculate the half-lives of the oA\_nc nuclei. The decomposed uncertainties are listed in Table \ref{tab:stat}. The systematic uncertainty is larger than $\sigma_{sys}(oA\_nc)$ for both formulas. Moreover, the residuals calculated by parameters obtained with ee nuclei are globally positive. It does indicate that the parameters obtained with ee nuclei underestimate the $\alpha$ decay half-lives of oA\_nc nuclei.
	
The second one is a systematic hindrance for $\alpha$ decay with the change of spin and/or parity. The half-lives of oA\_c nuclei are globally underestimated by the parameters from oA\_nc nuclei, as shown in Table \ref{tab:stat}. Although the parameters $a$ and $b$ of the oA\_nc nuclei drop in the 95\% confidence interval of the corresponding parameters of oA\_c nuclei. The parameter $c$ of the oA\_c nuclei presents a positive shift compared with the oA\_nc nuclei.

From the uncertainty analysis, both UDL and NGNL presents the largest statistical uncertainty and systematic uncertainty for the oA\_c nuclei. Particularly, $\sigma_{stat}(oA\_c)$ is three times larger than $\sigma_{stat}(oA\_nc)$, which indicates the weaker compatibility between these two formulas and oA\_c nuclei. Such a result coincides with the weakness of UDL. What confusing is that NGNL has accounted for the effect of angular momentum, but the uncertainties are as large as UDL, which should be due to the non-global property of the parameter $P$.

The third one is the value of parameter $P$. From the fitting results, one cannot rule out the possibility that the parameter $P$ takes zero or even negative values for parts of oA\_c nuclei, which contradicts the physical consideration that the change of spin and/or parity after decay should increase the $\alpha$ decay half-life. Actually, the value of $P$ strongly depends on the datasets used for fitting. It is difficult to obtain a global parameter $P$ based on the large differences among mean values of $P$ obtained from different datasets and corresponding large standard deviations, which are presented in Table \ref{tab:stat}.

	One recalls the change of spin in the $\alpha$ decay process for oA\_c nuclei compared to the oA\_nc nuclei. To investigate the existence of such a hindrance, the parameters obtained with oA\_nc nuclei are used to estimate the half-lives of oA\_c nuclei. Note that $P = 0$ in this extrapolation. The results are compared with the half-lives of oA\_c nuclei calculated with parameters obtained with itself. The residuals calculated with the parameters obtained with oA\_nc nuclei are generally positive, so the half-lives of oA\_c nuclei are underestimated. Moreover, an increment of systematic uncertainty comparing with $\sigma_{sys}(oA\_c)$ is noted, which are listed in Table \ref{tab:stat}. Hence, it proves the existence of a systematic hindrance in the $\alpha$ decay process of oA\_c nuclei. Residue of each nuclei is used to calculate the parameter $P$ together with angular momentum variation $l$. It is clearly seen in Table \ref{tab:stat} that the value of $P$ has a very large uncertainty, If only $l=2$ nuclei (around 65\% of all oA\_c nuclei) are considered, the uncertainty of $P$ is much reduced but remains to be large. The value of $P$ depends on the choices of different isotopes and isotones as the dataset, which is also reported by Ren \emph{et al.} in Ref.~\cite{ren2012new}.


	As the two types of residuals present an overall shift for decays from $l = 0$ to $l \ne 0$ relative to those of $l = 0$ to $l = 0$ decays. We, therefore, include a correction term for both UDL and NGNL, denoted as UDLC and NGNLC:
\begin{equation}
\begin{aligned}
	\log{T_{1/2}} = a_{1} Z_{c} Z_{d} \sqrt{\frac{A}{Q_{\alpha}}}+b_{1} \sqrt{A Z_{c} Z_{d} \left(A_{c}^{1 / 3}+A_{d}^{1 / 3} \right) } \\
	+ c_{1}  + 6P_{1} \delta
\end{aligned}
\end{equation}
\begin{equation}
	\log T_{1 / 2}=a_{2} Z_{c} Z_{d} \sqrt{\frac{A}{Q_{\alpha}}}+b_{2} \sqrt{A Z_{c} Z_{d}}+c_{2}+S+6P_{2}\delta
\end{equation}
where $\delta$ = 1 for oA\_c nuclei and $\delta$ = 0 for oA\_nc nuclei. Since 28 out of 43 channels in the oA\_c dataset are with $l=2$, $6=2(2+1)$ is set to be a coefficient of $P\delta$.

	The bootstrap is applied to analyze such two corrected formulas. The results are presented in Fig. \ref{fig:dis}, Table \ref{tab:stat} and Table \ref{pearson_oA}. In this case, parameters $P_{1}$ and $P_{2}$ stay positive, which is consistent with the physical consideration that the change of spin and parity hinder the $\alpha$ decay process. The relatively small uncertainties assure that the shift could be a global property. Moreover, observing that the systematic uncertainty is reduced while the statistical one is quasi invariant, it is believed that the constant correction is suitable to estimate the decay behavior.

%
%
%
%
\section{\label{summary}Summary}
	A framework based on the non-parametric bootstrap method is established in the present work to study the theoretical and statistical uncertainties of $\alpha$ decay models. The key objective is to reconstruct the distributions of several observables and to decompose the uncertainty for given theoretical models simultaneously. Independent from any specific assumption, it is quite generic and easy to implement with modern computational facilities.
	
	For simplicity, we only applied the model to study two existing empirical decay formulas for $\alpha$ decay half-life. The decomposed uncertainties also serve as an evaluator for the newly observed superallowed decay, $^{104}$Te $\rightarrow$ $^{100}$Sn. Our results suggest that this process is overestimated by the two laws with the parameters obtained from fitting to other nuclei while neglecting the experimental uncertainty on decay energy or theoretical uncertainty estimated in the present work. However, the observed half-life locates inside the estimated error bar when both the uncertainties are included. Therefore further measurements on the decay energy and half-life are recommended to assure the justification on superallowed nature.

Even if the same s-wave radiation of $\alpha$ is considered, uncertainty analysis clearly reveals that, as expected, oA nuclei have hindered the decay process comparing with neighboring ee nuclei due to the suppression of pairing. Moreover, oA nuclei with spin and/or parity change show additional hindrance compared with oA nuclei without spin or parity change, which can be related to changes in nuclear structure (or configuration) between the daughter and mother nuclei. Such hindrances deserve further investigations.
	
The present work emphasizes the analysis of the uncertainty of theoretical works. It reaffirms the reliability of conclusions from a parametric model. It is also straightforward to apply the model for a more complicated, non-linear case.

\section{\label{ack}Acknowledgements}
We thank Dr. Xiaoyun Li for useful discussions on the bootstrap method. This work has been supported by the National Natural Science Foundation
of China under Grant No.~11775316, the Tip-top Scientific and Technical Innovative Youth Talents of Guangdong special support program under Grant No.~2016TQ03N575, the National Key Research and Development Program of China under Grant No. 2018YFB1900405, and the computational resources from SYSU and National Supercomputer Center in Guangzhou. CQ is supported by the Swedish Research Council (VR) under grant Nos. 621-2012-3805, and 621-2013-4323 and the G\"{o}ran Gustafsson foundation.

\appendix*
\section{\label{appendix}Datasets}

Observed data of $\alpha$ decay used in the present work is listed in Table \ref{tab:app1}, \ref{tab:app2}, \ref{tab:app3} and \ref{tab:app4}.


\begin{longtable}{cccccc}
	\caption{Dataset I - Observed ground-state-to-ground-state $\alpha$ decay of even-even nuclei used in the present work.}
	\label{tab:app1} \\
	\hline
	Nucl.      & $E_{\alpha}$(keV) & $\sigma$ & Int.(\%) & $T_{1/2}$(s)         & $\sigma$             \\
	\hline
	\endfirsthead
	\multicolumn{6}{c}%
	{{ Table \thetable\ continued from previous page}} \\
	\hline
	Nucl.      & $E_{\alpha}$(keV) & $\sigma$ & Int.(\%) & $T_{1/2}$(s)         & $\sigma$             \\
	\hline
	\endhead
	\hline
	\endfoot
	\endlastfoot
	$^{106}$Te & 4290.22           & 9.35     & 100      & $7.80\times10^{-5}$  & $1.10\times10^{-05}$ \\
	$^{108}$Te & 3420.46           & 7.6      & 49       & $2.10\times10^{+00}$ & $1.00\times10^{-01}$ \\
	$^{112}$Xe & 3330.41           & 6.27     & 1.2      & $2.70\times10^{+00}$ & $8.00\times10^{-01}$ \\
	$^{114}$Ba & 3592.28           & 18.54    & 0.9      & $4.60\times10^{-01}$ & $1.25\times10^{-01}$ \\
	$^{144}$Nd & 1903.22           & 1.58     & 100      & $7.23\times10^{+22}$ & $5.05\times10^{+21}$ \\
	$^{146}$Sm & 2528.76           & 2.81     & 100      & $2.15\times10^{+15}$ & $2.21\times10^{+14}$ \\
	$^{148}$Sm & 1986.78           & 0.36     & 100      & $1.99\times10^{+23}$ & $4.10\times10^{+22}$ \\
	$^{148}$Gd & 3271.29           & 0.03     & 100      & $2.24\times10^{+09}$ & $3.16\times10^{+07}$ \\
	$^{150}$Gd & 2807.46           & 6.02     & 100      & $5.65\times10^{+13}$ & $2.52\times10^{+12}$ \\
	$^{152}$Gd & 2204.43           & 1.04     & 100      & $3.41\times10^{+21}$ & $2.52\times10^{+20}$ \\
	$^{150}$Dy & 4351.27           & 1.54     & 36       & $4.30\times10^{+02}$ & $3.00\times10^{+00}$ \\
	$^{152}$Dy & 3726.54           & 4.35     & 0.1      & $8.57\times10^{+03}$ & $7.20\times10^{+01}$ \\
	$^{154}$Dy & 2945.11           & 4.95     & 100      & $9.47\times10^{+13}$ & $4.73\times10^{+13}$ \\
	$^{152}$Er & 4934.26           & 1.62     & 90       & $1.03\times10^{+01}$ & $1.00\times10^{-01}$ \\
	$^{154}$Er & 4279.66           & 2.61     & 0.47     & $2.24\times10^{+02}$ & $5.40\times10^{+00}$ \\
	$^{156}$Er & 3481.25           & 25.06    & 0.00002  & $1.17\times10^{+03}$ & $6.00\times10^{+01}$ \\
	$^{154}$Yb & 5474.31           & 1.73     & 92.6     & $4.09\times10^{-01}$ & $2.00\times10^{-03}$ \\
	$^{156}$Yb & 4809.74           & 3.54     & 10       & $2.61\times10^{+01}$ & $7.00\times10^{-01}$ \\
	$^{158}$Yb & 4169.87           & 7.04     & 0.0021   & $8.94\times10^{+01}$ & $7.80\times10^{+00}$ \\
	$^{156}$Hf & 6028.57           & 3.81     & 97       & $2.30\times10^{-02}$ & $1.00\times10^{-03}$ \\
	$^{158}$Hf & 5404.78           & 2.72     & 44.3     & $9.90\times10^{-01}$ & $3.00\times10^{-02}$ \\
	$^{160}$Hf & 4901.86           & 2.59     & 0.7      & $1.36\times10^{+01}$ & $2.00\times10^{-01}$ \\
	$^{162}$Hf & 4416.31           & 4.86     & 0.008    & $3.94\times10^{+01}$ & $9.00\times10^{-01}$ \\
	$^{174}$Hf & 2494.52           & 2.26     & 100      & $6.31\times10^{+22}$ & $1.26\times10^{+22}$ \\
	$^{160}$W  & 6065.53           & 4.59     & 87       & $9.00\times10^{-02}$ & $5.00\times10^{-03}$ \\
	$^{162}$W  & 5678.27           & 2.4      & 45.2     & $1.19\times10^{+00}$ & $1.20\times10^{-01}$ \\
	$^{164}$W  & 5278.27           & 2.01     & 3.8      & $6.30\times10^{+00}$ & $2.00\times10^{-01}$ \\
	$^{166}$W  & 4856.06           & 3.94     & 0.035    & $1.92\times10^{+01}$ & $6.00\times10^{-01}$ \\
	$^{168}$W  & 4500.48           & 11.46    & 0.0032   & $5.09\times10^{+01}$ & $1.90\times10^{+00}$ \\
	$^{180}$W  & 2515.29           & 1.03     & 100      & $5.68\times10^{+25}$ & $6.31\times10^{+24}$ \\
	$^{166}$Os & 6142.77           & 3.32     & 72       & $2.13\times10^{-01}$ & $5.00\times10^{-03}$ \\
	$^{168}$Os & 5815.63           & 2.7      & 43       & $2.10\times10^{+00}$ & $1.00\times10^{-01}$ \\
	$^{170}$Os & 5536.87           & 2.69     & 9.5      & $7.37\times10^{+00}$ & $1.80\times10^{-01}$ \\
	$^{172}$Os & 5224.35           & 6.97     & 1.1      & $1.92\times10^{+01}$ & $9.00\times10^{-01}$ \\
	$^{174}$Os & 4870.5            & 9.69     & 0.024    & $4.40\times10^{+01}$ & $4.00\times10^{+00}$ \\
	$^{186}$Os & 2821.2            & 0.87     & 100      & $6.31\times10^{+22}$ & $3.47\times10^{+22}$ \\
	$^{168}$Pt & 6989.73           & 3.07     & 100      & $2.02\times10^{-03}$ & $1.00\times10^{-04}$ \\
	$^{172}$Pt & 6463.38           & 4.04     & 97       & $9.76\times10^{-02}$ & $1.30\times10^{-03}$ \\
	$^{174}$Pt & 6183.17           & 3.41     & 76       & $8.89\times10^{-01}$ & $1.70\times10^{-02}$ \\
	$^{176}$Pt & 5885.06           & 2.13     & 40       & $6.33\times10^{+00}$ & $1.50\times10^{-01}$ \\
	$^{178}$Pt & 5572.98           & 2.2      & 7.7      & $2.07\times10^{+01}$ & $7.00\times10^{-01}$ \\
	$^{180}$Pt & 5237.15           & 30.02    & 0.3      & $5.60\times10^{+01}$ & $3.00\times10^{+00}$ \\
	$^{182}$Pt & 4950.91           & 5.05     & 0.038    & $1.60\times10^{+02}$ & $7.20\times10^{+00}$ \\
	$^{184}$Pt & 4598.65           & 8.1      & 0.0017   & $1.04\times10^{+03}$ & $1.20\times10^{+01}$ \\
	$^{186}$Pt & 4319.75           & 18.16    & 0.00014  & $7.49\times10^{+03}$ & $1.80\times10^{+02}$ \\
	$^{188}$Pt & 4006.69           & 5.27     & 0.000026 & $8.81\times10^{+05}$ & $2.59\times10^{+04}$ \\
	$^{190}$Pt & 3268.58           & 0.59     & 100      & $2.05\times10^{+19}$ & $9.47\times10^{+17}$ \\
	$^{172}$Hg & 7523.8            & 6.19     & 100      & $2.31\times10^{-04}$ & $9.00\times10^{-06}$ \\
	$^{174}$Hg & 7233.27           & 6.01     & 100      & $2.00\times10^{-03}$ & $4.00\times10^{-04}$ \\
	$^{176}$Hg & 6897.05           & 5.55     & 90       & $2.03\times10^{-02}$ & $1.40\times10^{-03}$ \\
	$^{178}$Hg & 6577.35           & 2.98     & 89       & $2.67\times10^{-01}$ & $2.40\times10^{-03}$ \\
	$^{180}$Hg & 6258.5            & 2.37     & 48       & $2.59\times10^{+00}$ & $1.00\times10^{-02}$ \\
	$^{182}$Hg & 5995.68           & 4.65     & 13.8     & $1.08\times10^{+01}$ & $6.00\times10^{-02}$ \\
	$^{184}$Hg & 5662              & 4.38     & 1.11     & $3.09\times10^{+01}$ & $2.60\times10^{-01}$ \\
	$^{186}$Hg & 5204.44           & 9.89     & 0.016    & $8.28\times10^{+01}$ & $3.60\times10^{+00}$ \\
	$^{188}$Hg & 4707.41           & 15.65    & 0.000037 & $1.95\times10^{+02}$ & $9.00\times10^{+00}$ \\
	$^{178}$Pb & 7790.39           & 14.32    & 100      & $2.30\times10^{-04}$ & $1.50\times10^{-04}$ \\
	$^{180}$Pb & 7418.65           & 5.48     & 100      & $4.10\times10^{-03}$ & $3.00\times10^{-04}$ \\
	$^{184}$Pb & 6774.01           & 3.13     & 80       & $4.90\times10^{-01}$ & $2.50\times10^{-02}$ \\
	$^{186}$Pb & 6469.98           & 5.77     & 40       & $4.82\times10^{+00}$ & $3.00\times10^{-02}$ \\
	$^{188}$Pb & 6108.85           & 3.4      & 9.3      & $2.51\times10^{+01}$ & $1.00\times10^{-01}$ \\
	$^{190}$Pb & 5697.54           & 4.57     & 0.4      & $7.10\times10^{+01}$ & $1.00\times10^{+00}$ \\
	$^{192}$Pb & 5220.98           & 5.11     & 0.0059   & $2.10\times10^{+02}$ & $6.00\times10^{+00}$ \\
	$^{194}$Pb & 4737.84           & 16.72    & $7.3\times10^{-06}$  & $6.42\times10^{+02}$ & $3.60\times10^{+01}$ \\
	$^{210}$Pb & 3792.38           & 20.39    & $1.9\times10^{-06}$  & $7.01\times10^{+08}$ & $6.94\times10^{+06}$ \\
	$^{186}$Po & 8501.17           & 13.71    & 100      & $3.40\times10^{-05}$ & $1.20\times10^{-05}$ \\
	$^{190}$Po & 7693.27           & 7.22     & 100      & $2.46\times10^{-03}$ & $5.00\times10^{-05}$ \\
	$^{194}$Po & 6987.07           & 3.18     & 100      & $3.92\times10^{-01}$ & $4.00\times10^{-03}$ \\
	$^{196}$Po & 6658.06           & 2.41     & 98       & $5.56\times10^{+00}$ & $9.00\times10^{-02}$ \\
	$^{198}$Po & 6309.67           & 1.37     & 57       & $1.06\times10^{+02}$ & $1.44\times10^{+00}$ \\
	$^{200}$Po & 5981.64           & 1.85     & 11.1     & $6.91\times10^{+02}$ & $4.80\times10^{+00}$ \\
	$^{202}$Po & 5700.96           & 1.69     & 1.92     & $2.68\times10^{+03}$ & $2.40\times10^{+01}$ \\
	$^{204}$Po & 5484.89           & 1.37     & 0.67     & $1.27\times10^{+04}$ & $4.32\times10^{+01}$ \\
	$^{206}$Po & 5327.02           & 1.3      & 5.45     & $7.60\times10^{+05}$ & $8.64\times10^{+03}$ \\
	$^{208}$Po & 5215.38           & 1.3      & 100      & $9.15\times10^{+07}$ & $6.31\times10^{+04}$ \\
	$^{210}$Po & 5407.53           & 0.07     & 100      & $1.20\times10^{+07}$ & $1.73\times10^{+02}$ \\
	$^{212}$Po & 8954.2            & 0.11     & 100      & $2.95\times10^{-07}$ & $1.00\times10^{-09}$ \\
	$^{214}$Po & 7833.54           & 0.06     & 100      & $1.64\times10^{-04}$ & $2.70\times10^{-07}$ \\
	$^{216}$Po & 6906.35           & 0.51     & 100      & $1.45\times10^{-01}$ & $2.00\times10^{-03}$ \\
	$^{218}$Po & 6114.75           & 0.09     & 100      & $1.86\times10^{+02}$ & $7.20\times10^{-01}$ \\
	$^{194}$Rn & 7862.42           & 10.21    & 100      & $7.80\times10^{-04}$ & $1.60\times10^{-04}$ \\
	$^{196}$Rn & 7616.74           & 9.19     & 100      & $4.70\times10^{-03}$ & $1.10\times10^{-03}$ \\
	$^{200}$Rn & 7043.36           & 2.14     & 92       & $1.09\times10^{+00}$ & $1.60\times10^{-01}$ \\
	$^{202}$Rn & 6773.8            & 1.83     & 78       & $9.70\times10^{+00}$ & $1.00\times10^{-01}$ \\
	$^{204}$Rn & 6546.65           & 1.82     & 72.4     & $7.45\times10^{+01}$ & $1.38\times10^{+00}$ \\
	$^{206}$Rn & 6383.74           & 1.64     & 62       & $3.40\times10^{+02}$ & $1.02\times10^{+01}$ \\
	$^{208}$Rn & 6260.74           & 1.67     & 62       & $1.46\times10^{+03}$ & $8.40\times10^{+00}$ \\
	$^{210}$Rn & 6158.99           & 2.16     & 96       & $8.64\times10^{+03}$ & $3.60\times10^{+02}$ \\
	$^{212}$Rn & 6385.07           & 2.62     & 100      & $1.43\times10^{+03}$ & $7.20\times10^{+01}$ \\
	$^{214}$Rn & 9208.48           & 9.12     & 100      & $2.70\times10^{-07}$ & $2.00\times10^{-08}$ \\
	$^{216}$Rn & 8197.44           & 5.88     & 100      & $4.50\times10^{-05}$ & $5.00\times10^{-06}$ \\
	$^{218}$Rn & 7262.48           & 1.85     & 100      & $3.38\times10^{-02}$ & $1.50\times10^{-04}$ \\
	$^{220}$Rn & 6404.74           & 0.1      & 100      & $5.56\times10^{+01}$ & $1.00\times10^{-01}$ \\
	$^{222}$Rn & 5590.39           & 0.31     & 100      & $3.30\times10^{+05}$ & $1.73\times10^{+01}$ \\
	$^{202}$Ra & 7880.3            & 6.74     & 100      & $4.10\times10^{-03}$ & $1.10\times10^{-03}$ \\
	$^{204}$Ra & 7636.64           & 6.79     & 100      & $6.00\times10^{-02}$ & $9.00\times10^{-03}$ \\
	$^{208}$Ra & 7273.13           & 5.1      & 87       & $1.11\times10^{+00}$ & $4.50\times10^{-02}$ \\
	$^{214}$Ra & 7272.59           & 2.61     & 100      & $2.44\times10^{+00}$ & $1.60\times10^{-02}$ \\
	$^{216}$Ra & 9525.77           & 8.15     & 100      & $1.82\times10^{-07}$ & $1.00\times10^{-08}$ \\
	$^{218}$Ra & 8546              & 6.36     & 100      & $2.52\times10^{-05}$ & $3.00\times10^{-07}$ \\
	$^{220}$Ra & 7592.46           & 5.65     & 100      & $1.79\times10^{-02}$ & $1.40\times10^{-03}$ \\
	$^{222}$Ra & 6677.93           & 4.23     & 100      & $3.36\times10^{+01}$ & $4.00\times10^{-01}$ \\
	$^{224}$Ra & 5788.92           & 0.15     & 100      & $3.14\times10^{+05}$ & $1.99\times10^{+02}$ \\
	$^{226}$Ra & 4870.7            & 0.25     & 100      & $5.05\times10^{+10}$ & $2.21\times10^{+08}$ \\
	$^{208}$Th & 8202.03           & 30.59    & 100      & $2.40\times10^{-03}$ & $1.20\times10^{-03}$ \\
	$^{212}$Th & 7958.04           & 4.56     & 100      & $3.17\times10^{-02}$ & $1.30\times10^{-03}$ \\
	$^{214}$Th & 7827.18           & 5.4      & 100      & $8.70\times10^{-02}$ & $1.00\times10^{-02}$ \\
	$^{216}$Th & 8072.38           & 4.26     & 100      & $2.60\times10^{-02}$ & $2.00\times10^{-04}$ \\
	$^{218}$Th & 9849.09           & 9.11     & 100      & $1.17\times10^{-07}$ & $9.00\times10^{-09}$ \\
	$^{220}$Th & 8953.11           & 20.37    & 100      & $9.70\times10^{-06}$ & $6.00\times10^{-07}$ \\
	$^{222}$Th & 8127.03           & 5.09     & 100      & $2.24\times10^{-03}$ & $3.00\times10^{-05}$ \\
	$^{224}$Th & 7298.57           & 5.88     & 100      & $1.04\times10^{+00}$ & $2.00\times10^{-02}$ \\
	$^{226}$Th & 6452.53           & 1.01     & 100      & $1.84\times10^{+03}$ & $1.80\times10^{+00}$ \\
	$^{228}$Th & 5520.15           & 0.22     & 100      & $6.04\times10^{+07}$ & $2.52\times10^{+04}$ \\
	$^{230}$Th & 4769.85           & 1.52     & 100      & $2.38\times10^{+12}$ & $9.47\times10^{+09}$ \\
	$^{232}$Th & 4081.6            & 1.4      & 100      & $4.42\times10^{+17}$ & $3.16\times10^{+15}$ \\
	$^{216}$U  & 8530.63           & 26.21    & 100      & $6.90\times10^{-03}$ & $2.90\times10^{-03}$ \\
	$^{218}$U  & 8774.81           & 8.63     & 100      & $5.50\times10^{-04}$ & $1.40\times10^{-04}$ \\
	$^{222}$U  & 9481.17           & 50.92    & 100      & $4.70\times10^{-06}$ & $7.00\times10^{-07}$ \\
	$^{224}$U  & 8628.24           & 6.75     & 100      & $3.96\times10^{-04}$ & $1.70\times10^{-05}$ \\
	$^{226}$U  & 7700.84           & 4.27     & 100      & $2.69\times10^{-01}$ & $6.00\times10^{-03}$ \\
	$^{230}$U  & 5992.45           & 0.5      & 100      & $1.75\times10^{+06}$ & $1.73\times10^{+03}$ \\
	$^{232}$U  & 5413.63           & 0.09     & 100      & $2.17\times10^{+09}$ & $1.26\times10^{+07}$ \\
	$^{234}$U  & 4857.52           & 0.68     & 100      & $7.75\times10^{+12}$ & $1.89\times10^{+10}$ \\
	$^{236}$U  & 4572.94           & 0.87     & 100      & $7.39\times10^{+14}$ & $9.47\times10^{+11}$ \\
	$^{238}$U  & 4269.86           & 2.12     & 100      & $1.41\times10^{+17}$ & $1.89\times10^{+14}$ \\
	$^{228}$Pu & 7940.2            & 17.68    & 100      & $2.10\times10^{+00}$ & $1.30\times10^{+00}$ \\
	$^{230}$Pu & 7180.61           & 7.12     & 100      & $1.02\times10^{+02}$ & $1.02\times10^{+01}$ \\
	$^{232}$Pu & 6716              & 10.18    & 11       & $2.02\times10^{+03}$ & $3.00\times10^{+01}$ \\
	$^{234}$Pu & 6310.05           & 5.09     & 6        & $3.17\times10^{+04}$ & $3.60\times10^{+02}$ \\
	$^{236}$Pu & 5867.15           & 0.08     & 100      & $9.02\times10^{+07}$ & $2.52\times10^{+05}$ \\
	$^{238}$Pu & 5593.27           & 0.19     & 100      & $2.77\times10^{+09}$ & $3.16\times10^{+06}$ \\
	$^{240}$Pu & 5255.82           & 0.14     & 100      & $2.07\times10^{+11}$ & $2.21\times10^{+08}$ \\
	$^{242}$Pu & 4984.23           & 0.99     & 100      & $1.18\times10^{+13}$ & $6.31\times10^{+10}$ \\
	$^{244}$Pu & 4665.61           & 1.02     & 100      & $2.52\times10^{+15}$ & $2.84\times10^{+13}$ \\
	$^{234}$Cm & 7365.33           & 9.1      & 27       & $5.20\times10^{+01}$ & $9.00\times10^{+00}$ \\
	$^{236}$Cm & 7066.99           & 5.09     & 18       & $4.08\times10^{+02}$ & $4.80\times10^{+01}$ \\
	$^{238}$Cm & 6670.3            & 10.17    & 3.84     & $7.92\times10^{+03}$ & $1.44\times10^{+03}$ \\
	$^{240}$Cm & 6397.8            & 0.6      & 100      & $2.33\times10^{+06}$ & $8.64\times10^{+04}$ \\
	$^{242}$Cm & 6215.63           & 0.08     & 100      & $1.41\times10^{+07}$ & $1.73\times10^{+04}$ \\
	$^{244}$Cm & 5901.6            & 0.03     & 100      & $5.71\times10^{+08}$ & $6.31\times10^{+05}$ \\
	$^{246}$Cm & 5475.12           & 0.89     & 100      & $1.49\times10^{+11}$ & $1.26\times10^{+09}$ \\
	$^{248}$Cm & 5161.81           & 0.25     & 91.61    & $1.10\times10^{+13}$ & $1.89\times10^{+11}$ \\
	$^{240}$Cf & 7710.98           & 3.78     & 98.5     & $4.03\times10^{+01}$ & $9.00\times10^{-01}$ \\
	$^{242}$Cf & 7516.86           & 4.07     & 80       & $2.09\times10^{+02}$ & $9.00\times10^{+00}$ \\
	$^{244}$Cf & 7328.96           & 1.81     & 100      & $1.16\times10^{+03}$ & $3.60\times10^{+01}$ \\
	$^{246}$Cf & 6861.61           & 1        & 100      & $1.29\times10^{+05}$ & $1.80\times10^{+03}$ \\
	$^{248}$Cf & 6361.2            & 5        & 100      & $2.88\times10^{+07}$ & $2.42\times10^{+05}$ \\
	$^{250}$Cf & 6128.51           & 0.19     & 100      & $4.13\times10^{+08}$ & $2.84\times10^{+06}$ \\
	$^{252}$Cf & 6216.95           & 0.04     & 96.908   & $8.35\times10^{+07}$ & $2.52\times10^{+05}$ \\
	$^{254}$Cf & 5926.89           & 5.08     & 0.31     & $5.23\times10^{+06}$ & $1.73\times10^{+04}$ \\
	$^{248}$Fm & 7994.76           & 8.27     & 95       & $3.45\times10^{+01}$ & $1.20\times10^{+00}$ \\
	$^{252}$Fm & 7152.7            & 2        & 100      & $9.14\times10^{+04}$ & $1.44\times10^{+02}$ \\
	$^{254}$Fm & 7307.49           & 1.86     & 100      & $1.17\times10^{+04}$ & $7.20\times10^{+00}$ \\
	$^{256}$Fm & 7027.28           & 5.08     & 8.1      & $9.46\times10^{+03}$ & $7.80\times10^{+01}$ \\
	$^{254}$No & 8226.19           & 8.05     & 90       & $5.12\times10^{+01}$ & $4.00\times10^{-01}$ \\
	$^{256}$No & 8581.52           & 5.45     & 100      & $2.91\times10^{+00}$ & $5.00\times10^{-02}$ \\
	$^{256}$Rf & 8925.71           & 15.24    & 0.32     & $6.67\times10^{-03}$ & $1.00\times10^{-04}$ \\
	$^{258}$Rf & 9192.77           & 30.47    & 13       & $1.38\times10^{-02}$ & $9.00\times10^{-04}$ \\
	$^{260}$Sg & 9900.59           & 10.16    & 40       & $4.95\times10^{-03}$ & $3.30\times10^{-04}$ \\
	$^{264}$Hs & 10590.75          & 20.31    & 50       & $5.40\times10^{-04}$ & $3.00\times10^{-04}$ \\
	$^{270}$Ds & 11116.99          & 28.42    & 100      & $2.05\times10^{-04}$ & $4.80\times10^{-05}$ \\
	\hline
\end{longtable}


\begin{table*}[]
	\caption{Dataset II - Observed grand-state-to-excited-state $\alpha$ decay data of even-even nuclei used in the present work.}
	\label{tab:app2}
	\begin{ruledtabular}
	\begin{tabular}{cccccccc}
		Nucl.       & $E_{\alpha}$(keV) & $\sigma$ & Channel                       & $\Delta l$ & Int.(\%) & $T_{1/2}$(s)         & $\sigma$             \\
		\colrule
		$^{220}$Ra & 7131              & 6        & $ 0^{+} \rightarrow 2^{+} $   & 2          & 1        & $1.79\times10^{-02}$ & $1.40\times10^{-03}$ \\
		$^{222}$Ra & 6354.7            & 4        & $ 0^{+} \rightarrow 2^{+} $   & 2          & 3.05     & $3.36\times10^{+01}$ & $4.00\times10^{-01}$ \\
		$^{224}$Ra & 5547.87           & 0.15     & $ 0^{+} \rightarrow 2^{+} $   & 2          & 5.06     & $3.14\times10^{+05}$ & $1.99\times10^{+02}$ \\
		$^{226}$Ra & 4684.42           & 0.25     & $ 0^{+} \rightarrow 2^{+} $   & 2          & 6.16     & $5.05\times10^{+10}$ & $2.21\times10^{+08}$ \\
		$^{224}$Th & 7099.6            & 6        & $ 0^{+} \rightarrow 2^{+} $   & 2          & 19       & $1.04\times10^{+00}$ & $2.00\times10^{-02}$ \\
		$^{226}$Th & 6339.8            & 2.2      & $ 0^{+} \rightarrow 2^{+} $   & 2          & 22.8     & $1.84\times10^{+03}$ & $1.80\times10^{+00}$ \\
		$^{228}$Th & 5435.68           & 0.22     & $ 0^{+} \rightarrow 2^{+} $   & 2          & 26       & $6.04\times10^{+07}$ & $2.52\times10^{+04}$ \\
		$^{230}$Th & 4702.3            & 1.5      & $ 0^{+} \rightarrow 2^{+} $   & 2          & 23.4     & $2.38\times10^{+12}$ & $9.47\times10^{+09}$ \\
		$^{232}$Th & 4017.8            & 1.4      & $ 0^{+} \rightarrow 2^{+} $   & 2          & 21.7     & $4.42\times10^{+17}$ & $3.16\times10^{+15}$ \\
		$^{224}$U  & 8246.5            & 8        & $ 0^{+} \rightarrow 2^{+} $   & 2          & 3.4      & $3.96\times10^{-04}$ & $1.70\times10^{-05}$ \\
		$^{226}$U  & 7531.7            & 14       & $ 0^{+} \rightarrow 2^{+} $   & 2          & 15       & $2.69\times10^{-01}$ & $6.00\times10^{-03}$ \\
		$^{230}$U  & 5920.5            & 0.7      & $ 0^{+} \rightarrow 2^{+} $   & 2          & 32       & $1.75\times10^{+06}$ & $1.73\times10^{+03}$ \\
		$^{232}$U  & 5355.83           & 0.09     & $ 0^{+} \rightarrow 2^{+} $   & 2          & 31.55    & $2.17\times10^{+09}$ & $1.26\times10^{+07}$ \\
		$^{234}$U  & 4806.6            & 0.9      & $ 0^{+} \rightarrow 2^{+} $   & 2          & 28.42    & $7.75\times10^{+12}$ & $1.89\times10^{+10}$ \\
		$^{236}$U  & 4523.6            & 0.9      & $ 0^{+} \rightarrow 2^{+} $   & 2          & 26       & $7.39\times10^{+14}$ & $9.47\times10^{+11}$ \\
		$^{238}$U  & 4220.1            & 2.9      & $ 0^{+} \rightarrow 2^{+} $   & 2          & 21       & $1.41\times10^{+17}$ & $1.89\times10^{+14}$ \\
		$^{230}$Pu & 7121              & 8        & $ 0^{+} \rightarrow (2^{+}) $ & 2          & 19       & $1.02\times10^{+02}$ & $1.02\times10^{+01}$ \\
		$^{232}$Pu & 6657              & 10       & $ 0^{+} \rightarrow 2^{+} $   & 2          & 7.6      & $2.02\times10^{+03}$ & $3.00\times10^{+01}$ \\
		$^{234}$Pu & 6258.3            & 5        & $ 0^{+} \rightarrow 2^{+} $   & 2          & 1.9      & $3.17\times10^{+04}$ & $3.60\times10^{+02}$ \\
		$^{236}$Pu & 5819.47           & 0.08     & $ 0^{+} \rightarrow 2^{+} $   & 2          & 30.8     & $9.02\times10^{+07}$ & $2.52\times10^{+05}$ \\
		$^{238}$Pu & 5549.7            & 0.19     & $ 0^{+} \rightarrow 2^{+} $   & 2          & 28.98    & $2.77\times10^{+09}$ & $3.16\times10^{+06}$ \\
		$^{240}$Pu & 5210.55           & 0.14     & $ 0^{+} \rightarrow 2^{+} $   & 2          & 27.1     & $2.07\times10^{+11}$ & $2.21\times10^{+08}$ \\
		$^{242}$Pu & 4939.8            & 1        & $ 0^{+} \rightarrow 2^{+} $   & 2          & 23.4     & $1.18\times10^{+13}$ & $6.31\times10^{+10}$ \\
		$^{244}$Pu & 4621.5            & 1        & $ 0^{+} \rightarrow (2^{+}) $ & 2          & 19.4     & $2.52\times10^{+15}$ & $2.84\times10^{+13}$ \\
		$^{238}$Cm & 6574              & 40       & $ 0^{+} \rightarrow 2^{+} $   & 2          & 1.17     & $7.92\times10^{+03}$ & $1.44\times10^{+03}$ \\
		$^{240}$Cm & 6353.2            & 0.6      & $ 0^{+} \rightarrow 2^{+} $   & 2          & 28.8     & $2.33\times10^{+06}$ & $8.64\times10^{+04}$ \\
		$^{242}$Cm & 6171.46           & 0.08     & $ 0^{+} \rightarrow 2^{+} $   & 2          & 25.92    & $1.41\times10^{+07}$ & $1.73\times10^{+04}$ \\
		$^{244}$Cm & 5858.94           & 0.05     & $ 0^{+} \rightarrow 2^{+} $   & 2          & 23.1     & $5.71\times10^{+08}$ & $6.31\times10^{+05}$ \\
		$^{246}$Cm & 5430.3            & 1        & $ 0^{+} \rightarrow 2^{+} $   & 2          & 17.8     & $1.49\times10^{+11}$ & $1.26\times10^{+09}$ \\
		$^{248}$Cm & 5117.61           & 0.25     & $ 0^{+} \rightarrow 2^{+} $   & 2          & 16.52    & $1.10\times10^{+13}$ & $1.89\times10^{+11}$ \\
		$^{240}$Cf & 7674              & 10       & $ 0^{+} \rightarrow 2^{+} $   & 2          & 28       & $4.03\times10^{+01}$ & $9.00\times10^{-01}$ \\
		$^{244}$Cf & 7290.9            & 1.8      & $ 0^{+} \rightarrow (2^{+}) $ & 2          & 18       & $1.16\times10^{+03}$ & $3.60\times10^{+01}$ \\
		$^{246}$Cf & 6819.5            & 1        & $ 0^{+} \rightarrow 2^{+} $   & 2          & 20.6     & $1.29\times10^{+05}$ & $1.80\times10^{+03}$ \\
		$^{248}$Cf & 6318              & 5        & $ 0^{+} \rightarrow 2^{+} $   & 2          & 19.6     & $2.88\times10^{+07}$ & $2.42\times10^{+05}$ \\
		$^{250}$Cf & 6085.54           & 0.19     & $ 0^{+} \rightarrow 2^{+} $   & 2          & 17.11    & $4.13\times10^{+08}$ & $2.84\times10^{+06}$ \\
		$^{252}$Cf & 6173.47           & 0.04     & $ 0^{+} \rightarrow 2^{+} $   & 2          & 14.5     & $8.35\times10^{+07}$ & $2.52\times10^{+05}$ \\
		$^{248}$Fm & 7958              & 8        & $ 0^{+} \rightarrow 2^{+} $   & 2          & 19       & $3.45\times10^{+01}$ & $1.20\times10^{+00}$ \\
		$^{252}$Fm & 7111.2            & 2        & $ 0^{+} \rightarrow 2^{+} $   & 2          & 15       & $9.14\times10^{+04}$ & $1.44\times10^{+02}$ \\
		$^{254}$Fm & 7264.5            & 2        & $ 0^{+} \rightarrow 2^{+} $   & 2          & 14.2     & $1.17\times10^{+04}$ & $7.20\times10^{+00}$ \\
		$^{256}$Fm & 6981.3            & 5        & $ 0^{+} \rightarrow 2^{+} $   & 2          & 1.22     & $9.46\times10^{+03}$ & $7.80\times10^{+01}$ \\
		$^{256}$No & 8534.4            & 5        & $ 0^{+} \rightarrow 2^{+} $   & 2          & 12.9     & $2.91\times10^{+00}$ & $5.00\times10^{-02}$ \\
		$^{260}$Sg & 9850              & 10       & $ 0^{+} \rightarrow (2^{+}) $ & 2          & 5        & $4.95\times10^{-03}$ & $3.30\times10^{-04}$
	\end{tabular}
	\end{ruledtabular}
\end{table*}

\begin{table*}
\caption{\label{tab:app3}Dataset III - Observed $\alpha$ decay data of odd-A nuclei without spin or parity change used in the present work.}
	\begin{ruledtabular}
	\begin{tabular}{cccccccc}
		Nucl.      & $E_{\alpha}$(keV) & $\sigma$ & Channel                     & $\Delta l$ & Int.(\%) & $T_{1/2}$(s)         & $\sigma$             \\
		\colrule
		$^{147}$Sm & 2311              & 1        & $7/2^{-}\rightarrow7/2^{-}$ & 0          & 100      & $3.38\times10^{+18}$ & $3.16\times10^{+16}$ \\
		$^{163}$W  & 5519.49           & 50       & $7/2^{-}\rightarrow7/2^{-}$ & 0          & 14       & $2.63\times10^{+00}$ & $9.00\times10^{-02}$ \\
		$^{163}$Re & 6011.99           & 7.77     & $1/2^{+}\rightarrow1/2^{+}$ & 0          & 32       & $3.90\times10^{-01}$ & $7.00\times10^{-02}$ \\
		$^{167}$Os & 5980              & 50       & $7/2^{-}\rightarrow7/2^{-}$ & 0          & 50       & $8.39\times10^{-01}$ & $5.00\times10^{-03}$ \\
		$^{167}$Ir & 6504.89           & 2.64     & $1/2^{+}\rightarrow1/2^{+}$ & 0          & 43       & $2.93\times10^{-02}$ & $6.00\times10^{-04}$ \\
		$^{171}$Pt & 6610              & 50       & $7/2^{-}\rightarrow7/2^{-}$ & 0          & 98       & $4.40\times10^{-02}$ & $7.00\times10^{-03}$ \\
		$^{177}$Pt & 5643              & 3        & $5/2^{-}\rightarrow5/2^{-}$ & 0          & 5        & $1.06\times10^{+01}$ & $4.00\times10^{-01}$ \\
		$^{175}$Au & 6583              & 4        & $1/2^{+}\rightarrow1/2^{+}$ & 0          & 94       & $2.01\times10^{-01}$ & $3.00\times10^{-03}$ \\
		$^{183}$Hg & 6039              & 4        & $1/2^{-}\rightarrow1/2^{-}$ & 0          & 11.7     & $9.40\times10^{+00}$ & $7.00\times10^{-01}$ \\
		$^{185}$Hg & 5774              & 5        & $1/2^{-}\rightarrow1/2^{-}$ & 0          & 5.8      & $4.91\times10^{+01}$ & $1.00\times10^{+00}$ \\
		$^{213}$Po & 8536.1            & 2.6      & $9/2^{+}\rightarrow9/2^{+}$ & 0          & 100      & $3.72\times10^{-06}$ & $2.00\times10^{-08}$ \\
		$^{215}$Po & 7526.3 & 0.8 & $9/2^{+}\rightarrow9/2^{+}$ & 0 & 99.99977 & $1.78\times10^{-03}$ & $4.00\times10^{-06}$ \\
		$^{203}$At & 6210.1            & 0.8      & $9/2^{-}\rightarrow9/2^{-}$ & 0          & 27       & $4.44\times10^{+02}$ & $1.20\times10^{+01}$ \\
		$^{205}$At & 6019.5            & 1.7      & $9/2^{-}\rightarrow9/2^{-}$ & 0          & 10       & $1.61\times10^{+03}$ & $4.80\times10^{+01}$ \\
		$^{207}$At & 5872              & 3        & $9/2^{-}\rightarrow9/2^{-}$ & 0          & 8.6      & $6.48\times10^{+03}$ & $1.44\times10^{+02}$ \\
		$^{209}$At & 5757.1            & 2        & $9/2^{-}\rightarrow9/2^{-}$ & 0          & 4.1      & $1.95\times10^{+04}$ & $1.80\times10^{+02}$ \\
		$^{211}$At & 5982.4            & 1.3      & $9/2^{-}\rightarrow9/2^{-}$ & 0          & 41.8     & $2.60\times10^{+04}$ & $2.52\times10^{+01}$ \\
		$^{213}$At & 9254              & 5        & $9/2^{-}\rightarrow9/2^{-}$ & 0          & 100      & $1.25\times10^{-07}$ & $6.00\times10^{-09}$ \\
		$^{215}$At & 8178              & 4        & $9/2^{-}\rightarrow9/2^{-}$ & 0          & 99.95    & $1.00\times10^{-04}$ & $2.00\times10^{-05}$ \\
		$^{217}$At & 7201.3            & 1.2      & $9/2^{-}\rightarrow9/2^{-}$ & 0          & 99.89    & $3.23\times10^{-02}$ & $4.00\times10^{-04}$ \\
		$^{209}$Rn & 6155.5            & 2        & $5/2^{-}\rightarrow5/2^{-}$ & 0          & 16.9     & $1.73\times10^{+03}$ & $5.40\times10^{+01}$ \\
		$^{211}$Rn & 5896.9            & 1.4      & $1/2^{-}\rightarrow1/2^{-}$ & 0          & 17.3     & $5.26\times10^{+04}$ & $7.20\times10^{+02}$ \\
		$^{217}$Rn & 7887              & 3        & $9/2^{+}\rightarrow9/2^{+}$ & 0          & 100      & $5.40\times10^{-04}$ & $5.00\times10^{-05}$ \\
		$^{219}$Rn & 6544.3            & 0.3      & $5/2^{+}\rightarrow5/2^{+}$ & 0          & 7.5      & $3.96\times10^{+00}$ & $1.00\times10^{-02}$ \\
		$^{215}$Fr & 9540              & 7        & $9/2^{-}\rightarrow9/2^{-}$ & 0          & 100      & $8.60\times10^{-08}$ & $5.00\times10^{-09}$ \\
		$^{217}$Fr & 8469              & 4        & $9/2^{-}\rightarrow9/2^{-}$ & 0          & 100      & $1.90\times10^{-05}$ & $3.00\times10^{-06}$ \\
		$^{219}$Fr & 7448.5            & 1.8      & $9/2^{-}\rightarrow9/2^{-}$ & 0          & 98.8     & $2.00\times10^{-02}$ & $2.00\times10^{-03}$ \\
		$^{209}$Ra & 7140              & 10       & $5/2^{-}\rightarrow5/2^{-}$ & 0          & 99.3     & $4.70\times10^{+00}$ & $2.00\times10^{-01}$ \\
		$^{213}$Ra & 6751              & 2.3      & $1/2^{-}\rightarrow1/2^{-}$ & 0          & 39       & $1.64\times10^{+02}$ & $3.00\times10^{+00}$ \\
		$^{221}$Ra & 6731.3            & 2        & $5/2^{+}\rightarrow5/2^{+}$ & 0          & 38       & $2.80\times10^{+01}$ & $2.00\times10^{+00}$ \\
		$^{223}$Ra & 5709.8            & 0.3      & $3/2^{+}\rightarrow3/2^{+}$ & 0          & 25.2     & $9.88\times10^{+05}$ & $4.32\times10^{+03}$ \\
		$^{219}$Ac & 8830              & 50       & $9/2^{-}\rightarrow9/2^{-}$ & 0          & 100      & $1.18\times10^{-05}$ & $1.50\times10^{-06}$ \\
		$^{229}$Th & 4931.8            & 1.2      & $5/2^{+}\rightarrow5/2^{+}$ & 0          & 56.2     & $2.32\times10^{+11}$ & $5.05\times10^{+09}$ \\
		$^{229}$Th & 4988.3            & 1.2      & $5/2^{+}\rightarrow5/2^{+}$ & 0          & 10.2     & $2.32\times10^{+11}$ & $5.05\times10^{+09}$ \\
		$^{229}$Th & 5142.7            & 1.2      & $5/2^{+}\rightarrow5/2^{+}$ & 0          & 6.6      & $2.32\times10^{+11}$ & $5.05\times10^{+09}$ \\
		$^{219}$Pa & 10080             & 50       & $9/2^{-}\rightarrow9/2^{-}$ & 0          & 100      & $5.30\times10^{-08}$ & $1.00\times10^{-08}$ \\
		$^{221}$Pa & 9250              & 50       & $9/2^{-}\rightarrow9/2^{-}$ & 0          & 100      & $5.90\times10^{-06}$ & $1.70\times10^{-06}$ \\
		$^{231}$Pa & 4820              & 0.8      & $3/2^{-}\rightarrow3/2^{-}$ & 0          & 8.4      & $2.87\times10^{+11}$ & $9.64\times10^{+08}$ \\
		$^{231}$Pa & 5150              & 0.8      & $3/2^{-}\rightarrow3/2^{-}$ & 0          & 11       & $2.87\times10^{+11}$ & $9.64\times10^{+08}$ \\
		$^{233}$U  & 4908.5            & 1.2      & $5/2^{+}\rightarrow5/2^{+}$ & 0          & 84.3     & $1.40\times10^{+12}$ & $1.75\times10^{+09}$ \\
		$^{235}$U  & 4290.4            & 0.7      & $7/2^{-}\rightarrow7/2^{-}$ & 0          & 6.01     & $2.22\times10^{+16}$ & $1.58\times10^{+13}$ \\
		$^{237}$Np & 4720.4            & 1.2      & $5/2^{+}\rightarrow5/2^{+}$ & 0          & 6.43     & $1.88\times10^{+13}$ & $6.14\times10^{+10}$ \\
		$^{237}$Np & 4746              & 1.2      & $5/2^{+}\rightarrow5/2^{+}$ & 0          & 3.478    & $1.88\times10^{+13}$ & $6.14\times10^{+10}$ \\
		$^{237}$Np & 4871.8            & 1.2      & $5/2^{+}\rightarrow5/2^{+}$ & 0          & 47.64    & $1.88\times10^{+13}$ & $6.14\times10^{+10}$ \\
		$^{239}$Pu & 5224.4            & 0.21     & $1/2^{+}\rightarrow1/2^{+}$ & 0          & 70.77    & $2.11\times10^{+11}$ & $2.63\times10^{+08}$ \\
		$^{241}$Am & 5578.32           & 0.12     & $5/2^{-}\rightarrow5/2^{-}$ & 0          & 84.8     & $1.37\times10^{+10}$ & $1.89\times10^{+07}$ \\
		$^{245}$Cf & 7258.5            & 1.8      & $1/2^{+}\rightarrow1/2^{+}$ & 0          & 32.4     & $2.70\times10^{+03}$ & $7.80\times10^{+01}$ \\
		$^{253}$Es & 6739.24           & 0.05     & $7/2^{+}\rightarrow7/2^{+}$ & 0          & 100      & $1.77\times10^{+06}$ & $2.59\times10^{+03}$ \\
		$^{255}$Fm & 7133.4            & 1.8      & $7/2^{+}\rightarrow7/2^{+}$ & 0          & 93.4     & $7.23\times10^{+04}$ & $2.52\times10^{+02}$
	\end{tabular}
	\end{ruledtabular}
\end{table*}

\begin{table*}
\caption{\label{tab:app4}Dataset IV - Observed $\alpha$ decay data of odd-A nuclei with spin and/or parity change used in the present work.}
	\begin{ruledtabular}
	\begin{tabular}{cccccccc}
		Nucl.      & $E_{\alpha}$(keV) & $\sigma$ & Channel                      & $\Delta l$ & Int.(\%) & $T_{1/2}$(s)         & $\sigma$             \\
		\colrule
		$^{149}$Tb & 4077.5            & 2.2      & $1/2^{+}\rightarrow5/2^{+}$  & 2          & 16.7     & $1.48\times10^{+04}$ & $9.00\times10^{+01}$ \\
		$^{209}$Bi & 3137.2            & 0.8      & $9/2^{-}\rightarrow1/2^{+}$  & 5          & 99.92    & $6.00\times10^{+26}$ & $6.31\times10^{+25}$ \\
		$^{211}$Bi & 6399.2            & 0.5      & $9/2^{-}\rightarrow3/2^{+}$  & 3          & 16.19    & $1.28\times10^{+02}$ & $1.20\times10^{+00}$ \\
		$^{211}$Bi & 6750.3            & 0.5      & $9/2^{-}\rightarrow1/2^{+}$  & 5          & 83.54    & $1.28\times10^{+02}$ & $1.20\times10^{+00}$ \\
		$^{213}$Bi & 5988              & 4        & $9/2^{-}\rightarrow1/2^{+}$  & 5          & 1.959    & $2.74\times10^{+03}$ & $2.40\times10^{+00}$ \\
		$^{211}$Rn & 5965.4            & 1.4      & $1/2^{-}\rightarrow5/2^{-}$  & 2          & 9.3      & $5.26\times10^{+04}$ & $7.20\times10^{+02}$ \\
		$^{219}$Rn & 6674.9            & 0.3      & $5/2^{+}\rightarrow7/2^{+}$  & 2          & 12.9     & $3.96\times10^{+00}$ & $1.00\times10^{-02}$ \\
		$^{219}$Rn & 6946.1            & 0.3      & $5/2^{+}\rightarrow9/2^{+}$  & 2          & 79.4     & $3.96\times10^{+00}$ & $1.00\times10^{-02}$ \\
		$^{213}$Ra & 6646.4            & 2.3      & $1/2^{-}\rightarrow3/2^{-}$  & 2          & 4.6      & $1.64\times10^{+02}$ & $3.00\times10^{+00}$ \\
		$^{213}$Ra & 6861.3            & 2.3      & $1/2^{-}\rightarrow5/2^{-}$  & 2          & 36       & $1.64\times10^{+02}$ & $3.00\times10^{+00}$ \\
		$^{221}$Ra & 6880.4            & 2        & $5/2^{+}\rightarrow9/2^{+}$  & 2          & 32       & $2.80\times10^{+01}$ & $2.00\times10^{+00}$ \\
		$^{223}$Ra & 5979.3            & 0.3      & $3/2^{+}\rightarrow5/2^{+}$  & 2          & 1        & $9.88\times10^{+05}$ & $4.32\times10^{+03}$ \\
		$^{229}$Th & 4883.6            & 1.2      & $5/2^{+}\rightarrow7/2^{+}$  & 2          & 1.5      & $2.32\times10^{+11}$ & $5.05\times10^{+09}$ \\
		$^{229}$Th & 5067.6            & 1.2      & $5/2^{+}\rightarrow9/2^{+}$  & 2          & 3.17     & $2.32\times10^{+11}$ & $5.05\times10^{+09}$ \\
		$^{229}$Th & 4900.2            & 1.2      & $5/2^{+}\rightarrow7/2^{+}$  & 2          & 9.3      & $2.32\times10^{+11}$ & $5.05\times10^{+09}$ \\
		$^{229}$Th & 4924.5            & 1.2      & $5/2^{+}\rightarrow7/2^{+}$  & 2          & 5        & $2.32\times10^{+11}$ & $5.05\times10^{+09}$ \\
		$^{229}$Th & 5056.5            & 1.2      & $5/2^{+}\rightarrow7/2^{+}$  & 2          & 5.97     & $2.32\times10^{+11}$ & $5.05\times10^{+09}$ \\
		$^{231}$Pa & 4762.8            & 0.8      & $3/2^{-}\rightarrow7/2^{-}$  & 2          & 1.5      & $1.03\times10^{+12}$ & $3.47\times10^{+09}$ \\
		$^{231}$Pa & 4795.5            & 0.8      & $3/2^{-}\rightarrow1/2^{-}$  & 2          & 1.268    & $1.03\times10^{+12}$ & $3.47\times10^{+09}$ \\
		$^{231}$Pa & 4939.2            & 0.8      & $3/2^{-}\rightarrow13/2^{+}$ & 5          & 1.4      & $1.03\times10^{+12}$ & $3.47\times10^{+09}$ \\
		$^{231}$Pa & 5040              & 0.8      & $3/2^{-}\rightarrow9/2^{+}$  & 3          & 22.8     & $1.03\times10^{+12}$ & $3.47\times10^{+09}$ \\
		$^{231}$Pa & 5075.9            & 0.8      & $3/2^{-}\rightarrow7/2^{-}$  & 2          & 1.4      & $1.03\times10^{+12}$ & $3.47\times10^{+09}$ \\
		$^{231}$Pa & 5122.6            & 0.8      & $3/2^{-}\rightarrow3/2^{+}$  & 1          & 2.5      & $1.03\times10^{+12}$ & $3.47\times10^{+09}$ \\
		$^{231}$Pa & 5023.2            & 0.8      & $3/2^{-}\rightarrow9/2^{-}$  & 4          & 3        & $1.03\times10^{+12}$ & $3.47\times10^{+09}$ \\
		$^{231}$Pa & 5120              & 0.8      & $3/2^{-}\rightarrow5/2^{-}$  & 2          & 20       & $1.03\times10^{+12}$ & $3.47\times10^{+09}$ \\
		$^{231}$Pa & 5103.7            & 0.8      & $3/2^{-}\rightarrow5/2^{+}$  & 1          & 25.4     & $1.03\times10^{+12}$ & $3.47\times10^{+09}$ \\
		$^{233}$U  & 4811.4            & 1.2      & $5/2^{+}\rightarrow9/2^{+}$  & 2          & 1.61     & $5.02\times10^{+12}$ & $6.31\times10^{+09}$ \\
		$^{233}$U  & 4866.1            & 1.2      & $5/2^{+}\rightarrow7/2^{+}$  & 2          & 13.2     & $5.02\times10^{+12}$ & $6.31\times10^{+09}$ \\
		$^{235}$U  & 4492.5            & 0.7      & $7/2^{-}\rightarrow5/2^{-}$  & 2          & 3.09     & $2.22\times10^{+16}$ & $1.58\times10^{+13}$ \\
		$^{235}$U  & 4582.1            & 0.7      & $7/2^{-}\rightarrow9/2^{+}$  & 1          & 1.28     & $2.22\times10^{+16}$ & $1.58\times10^{+13}$ \\
		$^{235}$U  & 4636.2            & 0.7      & $7/2^{-}\rightarrow7/2^{+}$  & 1          & 3.82     & $2.22\times10^{+16}$ & $1.58\times10^{+13}$ \\
		$^{235}$U  & 4678.2            & 0.7      & $7/2^{-}\rightarrow5/2^{+}$  & 1          & 4.77     & $2.22\times10^{+16}$ & $1.58\times10^{+13}$ \\
		$^{235}$U  & 4441.3            & 0.7      & $7/2^{-}\rightarrow9/2^{-}$  & 2          & 18.92    & $2.22\times10^{+16}$ & $1.58\times10^{+13}$ \\
		$^{237}$Np & 4854.7            & 1.2      & $5/2^{+}\rightarrow7/2^{+}$  & 2          & 23.2     & $6.77\times10^{+13}$ & $2.21\times10^{+11}$ \\
		$^{237}$Np & 4887.8            & 1.2      & $5/2^{+}\rightarrow5/2^{-}$  & 1          & 2.014    & $6.77\times10^{+13}$ & $2.21\times10^{+11}$ \\
		$^{237}$Np & 4901.2            & 1.2      & $5/2^{+}\rightarrow7/2^{-}$  & 1          & 2.43     & $6.77\times10^{+13}$ & $2.21\times10^{+11}$ \\
		$^{237}$Np & 4958.3            & 1.2      & $5/2^{+}\rightarrow3/2^{-}$  & 1          & 2.39     & $6.77\times10^{+13}$ & $2.21\times10^{+11}$ \\
		$^{237}$Np & 4849.3            & 1.2      & $5/2^{+}\rightarrow9/2^{+}$  & 2          & 9.3      & $6.77\times10^{+13}$ & $2.21\times10^{+11}$ \\
		$^{239}$Pu & 5192.8            & 0.21     & $1/2^{+}\rightarrow5/2^{+}$  & 2          & 11.94    & $7.61\times10^{+11}$ & $9.47\times10^{+08}$ \\
		$^{239}$Pu & 5231.5            & 0.21     & $1/2^{+}\rightarrow3/2^{+}$  & 2          & 17.11    & $7.61\times10^{+11}$ & $9.47\times10^{+08}$ \\
		$^{241}$Am & 5479.32           & 0.12     & $5/2^{-}\rightarrow9/2^{-}$  & 2          & 1.66     & $1.37\times10^{+10}$ & $1.89\times10^{+07}$ \\
		$^{241}$Am & 5534.82           & 0.12     & $5/2^{-}\rightarrow7/2^{-}$  & 2          & 13.1     & $1.37\times10^{+10}$ & $1.89\times10^{+07}$ \\
		$^{255}$Fm & 7073.4            & 1.8      & $7/2^{+}\rightarrow9/2^{+}$  & 2          & 5.04     & $7.23\times10^{+04}$ & $2.52\times10^{+02}$
	\end{tabular}
	\end{ruledtabular}
\end{table*}

\bibliography{apstemplate}

\end{document}